\documentclass[letterpaper,times]{IONconf-v2}
\usepackage[hidelinks]{hyperref}

\usepackage{amsmath,amssymb} 
\usepackage{xcolor}

\usepackage{xcolor}
\usepackage{tabularray}
\usepackage{booktabs}
\usepackage{multirow}

\articletype{Regular papers}%

\received{}
\revised{}
\accepted{}
\doi{10.1109/xxx.xx.xxxx}
\journalname{NAVIGATION}
\journalvolume{X}
\journalnumber{X}

\title{GNSS Jamming and Spoofing Monitoring Using Low-Cost COTS Receivers}

\author[1]{Argyris Kriezis}

\author[1]{Yu-Hsuan Chen}

\author[1,2]{Dennis Akos}

\author[1]{Sherman Lo}

\author[1]{Todd Walter}

\authormark{Kriezis A. \textsc{et al}}

\address[1]{Stanford University}

\address[2]{University of Colorado, Boulder}

\corres{Argyris Kriezis\\ \email{akriezis@stanford.edu}}

\begin{document}

\abstract[Abstract]{The Global Navigation Satellite System (GNSS) is increasingly vulnerable to radio frequency interference (RFI), including jamming and spoofing, which threaten the integrity and availability of navigation and timing services. This paper presents a methodology for detecting and classifying RFI events using low-cost commercial off-the-shelf (COTS) GNSS receivers. By combining carrier-to-noise ratio (C/N\textsubscript{0}) measurements of SBAS and GPS satellites with a calibrated spectrum adjusted received power metric, a two-dimensional detection space is constructed to identify and distinguish nominal, jammed and spoofed signal conditions. The method is validated with data from Jammertest in Norway which offers controlled but realistic RFI conditions. Results demonstrate that COTS-based RFI detection, when properly processed, offers a viable and effective approach for GNSS RFI monitoring.}


\maketitle

\section{Introduction}

Global Navigation Satellite System (GNSS) signals are transmitted by satellites 20,000+ km above the Earth's surface and arrive at the ground, with power levels below that of thermal noise, making them highly susceptible to Radio Frequency Interference (RFI). RFI can manifest as jamming, where a source emits enough unwanted energy to prevent GNSS signal processing, or as spoofing, where an adversary transmits deceptive GNSS-like signals to mislead the receiver into calculating an incorrect position/time. In both cases, the impact on GNSS users can be substantial, underscoring the importance of monitoring the RFI environment within the Radio Navigation Satellite Service (RNSS) bands and detecting both intentional and unintentional sources of disruption.  

Prior to 2023, comparatively few RFI events affecting aviation were reported, with the ones in Denver Airport, Dallas Airport and Nantes, France Airport being among the most notable ones \citep{goward2023denver} \citep{liu2023investigation} \citep{rntf2017gpsjammer}. While EUROCONTROL has recorded GNSS outages from 2014, since 2022, the share of flights reporting GNSS RFI events has rapidly increased with many affected areas spreading from northeastern Europe to the Mediterranean Sea \citep{eurocontrol2021rfi} \citep{felux2024gnssjamming}. RFI hotspots have also been recorded in other parts of the world, such as Myanmar, the South China Sea, and the India-Pakistan border \citep{Lo2026ADSB}. The common aspect of nearly all of these regions is their proximity to conflict zones. The International Telecommunications Union (ITU), International Civil Aviation Organization (ICAO) and International Maritime Organization (IMO), the global organizations for spectrum, aviation and maritime safety, released a join statement in April, 2025 highlighting the need for members states to "take necessary measures to prevent satellite systems from suffering harmful interference" \citep{itu2025rnss}. However, none of these organizations has an enforcement mechanism that can force a nation to halt GNSS jamming and spoofing, especially when national security is on the line. As a result, it is important to build technologies that allow users to monitor, detect, and possibly isolate the impacts of RFI.    

\subsection{RFI Monitoring Techniques}
GNSS RFI monitoring techniques have been a study topic for years, with methods broadly categorized as global or local. Global methods offer wide-area geographic coverage, while local approaches are designed to monitor specific assets, such as airports, ports or highways. Satellite-based RFI monitoring has been shown to provide reliable detection and localization of interference sources with global Earth coverage \citep{clements2023, Sarda2018HawkEye360}. However, offering continuous interference monitoring even in small geographical regions requires a high number of satellites, making this solution costly to scale. The lower cost alternative to wide-scale RFI monitoring is the aggregation and processing of data from the Automatic Dependent Surveillance–Broadcast (ADS-B) exchange or the International GNSS Service (IGS) network. ADS-B data are transmitted by aircraft and include a GNSS signal quality indicator that can be used as a proxy for the interference. Multiple online platforms, such as \url{rfi.stanford.edu} and \url{gpsjam.org}, have demonstrated that ADS-B data can effectively highlight areas of RFI activity. However, the GNSS signal-quality metrics in ADS-B data also flags ionospheric scintillation and poor satellite visibility as interference leading to false-positive detections \citep{liu2022gnss}. Data from the IGS network, while still susceptible to false detections caused by environmental effects, have been successfully used to detect RFI events across the United States \citep{Jadanavi26}. Both ADS-B and IGS data are limited by their spatial resolution. ADS-B data are only available in areas with well monitored and dense air traffic. The sparse distribution of IGS ground stations limits the ability to itentify the RFI sources or assess the spatial extent of an RFI event.         

Local RFI monitoring systems have been deployed across airports in Europe, with Enaire's DYLEMA, deployed at Madrid Airport, being an example \citep{madridcobos2025gnss}. The system comprises nine stations designed for jamming and spoofing detection, most of which are equipped with multi-beam angle-of-arrival (AOA) antennas. It is capable of detecting and localizing RFI events in and around the airport. However, the system has two main drawbacks. First, the stations are positioned along the airport perimeter, which may result in missed detections of low-power RFI sources located within the airport grounds that affect only a small radius. Second, the system’s cost, on the order of millions of euros, which poses a significant barrier to scalability, particularly in countries like the US with hundreds of regional airports.

An alternative RFI monitoring system is being developed by the Stanford GPS Lab, aiming to strike a balance between detection sensitivity and affordability. The system leverages observables from inexpensive u-blox F9P COTS GNSS receivers to detect and classify interference events. While these receivers lack the data availability and accuracy of a spectrum analyzer or multi-antenna system, their sub \$400 cost allows for deployment at scale. Instead of placing a few detectors at an airport's perimeter, dozens of receivers can be spread out throughout an airport, port, or other area containing critical infrastructure. A large network of receivers provides spatial diversity, enhancing the ability to detect even low-power RFI sources within the airport environment.

\subsection{RFI Detection Metrics}

Detection of RFI events primarily relies on either direct observation of the RF environment or monitoring the integrity and availability of GNSS signals. FFT-based jamming detection using a spectrum analyzer or software-defined radio is a common approach for identifying increases in noise \citep{Moussa2017}. However, the cost of spectrum analyzers and high-quality software-defined radios has led researchers to use RF noise proxies available in COTS GNSS receivers, such as the Automatic Gain Control (AGC) observable \citep{bastide2003agc, Miralles2020SpoofingDetection}. In addition to AGC, the signal C/N\textsubscript{0} has been widely used to detect jamming by monitoring its degradation relative to nominal values \citep{Clements2025, Jadanavi26}.

A key limitation of both AGC and C/N\textsubscript{0} is their saturation at relatively low jamming powers. Although they remain effective for detecting the presence of jamming, they provide limited information about the interference power level or type once saturation occurs. Furthermore, the fidelity of AGC measurements depends on receiver quality. Low-cost receivers often provide only coarse gain quantization and apply AGC across the entire L1 band, making them sensitive to interference adjacent to GPS signals.  

In previously published work, the Stanford GPS lab has surveyed the observables available in the u-blox F9P receivers to determine the most effective ones in detecting jamming \citep{Kriezis2024, Kriezis2024b}. In that work, the u-blox SPAN (signal characteristics) message was identified, which provides a Fast Fourier
Transform (FFT) of the received signal with a frequency resolution of 500 kHz. The SPAN message information was used to determine the relative jamming level of RFI events and their spectral density. This paper extends this work, developing a processing and calibration methodology that allows for the conversion of the SPAN message information into the spectrally adjusted power density for the GPS L1 C/A signal.  

Spoofing detection typically relies on signal integrity metrics that evaluate the consistency of pseudorange measurements, Doppler, and receiver clock bias \citep{Jiaxun2018, Hwang2014, Liu2018}. However, spoofing can take many forms, making it difficult to rely on a single class of metrics for robust detection \citep{Psiaki2016}. While targeted spoofing attacks may require sophisticated defenses involving multiple antennas and signal integrity monitoring, most observed real-world spoofing events are non-targeted, wide-area attacks \citep{lo2025spoofing}. In prior work, data from a real-world spoofing event in the southeastern Mediterranean Sea were used to evaluate the effectiveness of spoofing detection metrics available in COTS GNSS receivers \citep{kriezis2025spoofing}. Among these metrics, signal C/N\textsubscript{0} proved particularly effective because, in wide-area spoofing scenarios, the spoofed signal power cannot match the expected GPS signal power at all locations. Hence, the imbalance between the expected and measured power can be leveraged to identify spoofed signals. 

Modeling the relationship between the expected signal C/N\textsubscript{0} and the noise level further improves spoofing detection sensitivity by constraining the power levels that a spoofing attack must target. It also enables clasification of attacks that combine spoofing and jamming regardless of their power level. Prior work has explored the relationship between C/N\textsubscript{0} and Automatic Gain Control (AGC) measurements for spoofing detection \citep{lo2021practical, Strizic2018, Miralles2020SpoofingDetection}. However, the effectiveness of these methods is limited by the coarse gain quantization of AGC measurements and the large nominal variation in C/N\textsubscript{0} values from non-geostationary orbit (NGSO) satellite signals. This paper improves spoofing classification method by replacing the AGC metric with the more accurate spectrum-adjusted received power density and utilizing SBAS signals, which provide more stable C/N\textsubscript{0} values, in addition to those from GPS.             

For an RFI monitoring system to be effective, it must balance detection sensitivity and robustness. This requires selecting appropriate thresholds for both RFI detection and classification between jamming and spoofing. No universally accepted threshold exists for defining jamming, and receiver manufacturers often use different criteria for their jamming flags. RTCA/EUROCAE's Minimum Operational Performance Standard (MOPS) provides a reference by defining the maximum interference level at which an aviation receiver must continue to operate \citep{eurocae_ed259b}. However, a monitoring station's detection threshold must be more sensitive to ensure that low-power RFI events are recorded. The threshold setting technique presented is designed to maximize a system's sensitivity while maintaining a nearly zero false positive rate. Additionally, it can autonomously adjust the thresholds when a receiver is deployed in a new environment.      

This paper makes four main contributions. First, it develops the spectrum-adjusted received power density metric for accurate RFI detection and characterization using only observables available from the low-cost u-blox F9P receiver. Second, it improves spoofing classification through the use of SBAS signals alongside the developed received power, compared to AGC-based methods. Third, it introduces a threshold-setting methodology that does not require manual calibration for deployment at new locations. Finally, it validates the proposed COTS receiver-based RFI monitoring method under real-world open-sky conditions during the 2024 Jammertest campaign in Norway.

\section{RFI Metrics and Receiver Calibration}
The u-blox F9P receiver provides a low-cost and scalable platform, however, it introduces two key challenges. First, the number and quality of the available observables is limited. Second, the hardware is susceptible to environmental effects such as temperature. As a result, the observables must be calibrated and processed to achieve reliable RFI detection and classification. This section presents the two selected metrics, C/N\textsubscript{0} and received power, and the methodology used to derive them from the available u-blox observables. These metrics are then combined to form a two-dimensional C/N\textsubscript{0} over received-power metric, which enables both interference detection and classification into jamming or spoofing. 

\subsection{C/N\textsubscript{0} Metric}
The C/N\textsubscript{0} metric is one of the most widely used indicators for both RFI detection and assessment of GNSS signal tracking quality prior to its use in a position solution \citep{sanmiguel2023calibration}. Lacking a standardized minimum threshold, this study establishes 27 dB-Hz as the baseline for a healthy signal. As shown in Equation \ref{eq:cn0}, the carrier component of the C/N\textsubscript{0} ratio depends on the satellite transmit power ($P_{tx}$), satellite antenna gain ($G_{tx}$), propagation loss ($L_{prop}$), receiver antenna gain ($G_{rx}$), and other losses ($L_{rem}$), including multipath, antenna, and atmospheric losses. The noise density component ($N_{0}$) is relatively stable at a given location and consists of the receiver noise floor, noise contributions from all visible GNSS satellites, emissions from systems operating within the GNSS band such as Distance Measuring Equipment (DME), and out-of-band emissions from systems operating in adjacent frequency bands.

\begin{equation}
\frac{C}{N_{0}} = \frac{P_{tx}+G_{tx}-L_{prop}+G_{Rx}-L_{rem}}{N_{0}}
\label{eq:cn0}
\end{equation}

For NGSO GNSS satellites, the transmit power typically remains constant. However, as the satellite moves across the sky, the transmit gain, propagation loss, and receive gain vary due to changes in both the satellite-user distance, the satellite elevation angle and multipath. Consequently, significant variation in C/N\textsubscript{0} is observed, even after calibrating the measurements with the elevation angle.

\begin{figure}[htb]
\centering
\includegraphics[scale=0.26]{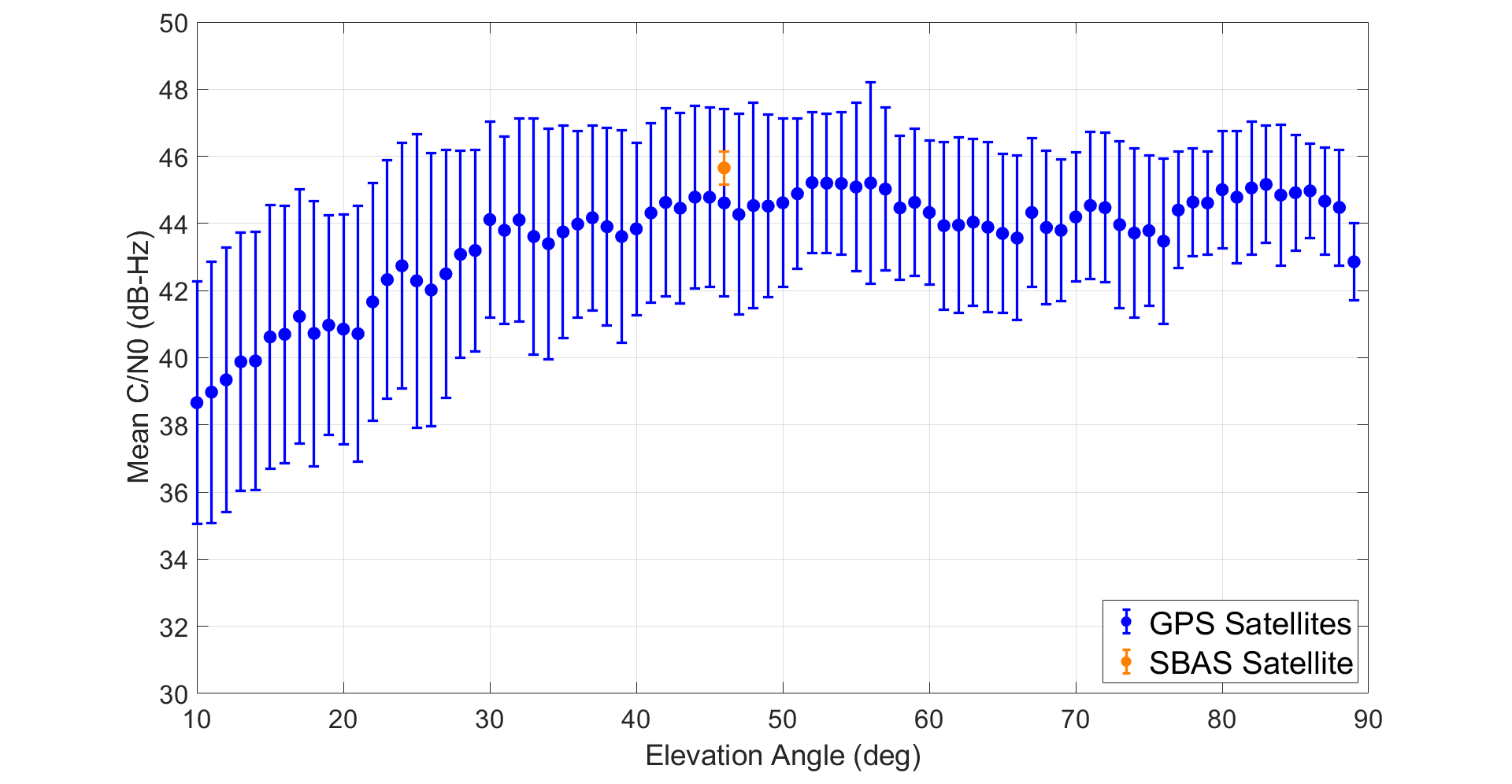}
\caption{Mean GPS L1 C/N\textsubscript{0} versus elevation angle computed from 24 hours of RFI-free measurements collected at Stanford, CA}
\label{fig:NAV25_CN0_Elev}
\end{figure}

In contrast, the geostationary (GEO) satellites of the Space-Based Augmentation System (SBAS) maintain a fixed position in the sky relative to a user on Earth. As a result, their C/N\textsubscript{0} values exhibit substantially less temporal variation than those of NGSO GPS and Galileo satellites. Figure~\ref{fig:NAV25_CN0_Elev} presents the mean C/N\textsubscript{0} values and corresponding standard deviations for GPS and SBAS satellites observed from Stanford University's campus under nominal RFI-free conditions. The figure highlights the significantly lower C/N\textsubscript{0} variability of SBAS signals. 

Since the carrier component of C/N\textsubscript{0} can be estimated, the noise component can be used as an indicator of jamming. As shown in Equation \ref{eq:cn0_jam}, C/N\textsubscript{0} decreases when additional interference noise is introduced, where $I$ denotes the spectrum-adjusted jamming power density. Although this relationship applies to both GPS and SBAS signals, the lower variability of SBAS C/N\textsubscript{0} measurements permits tighter detection thresholds, resulting in improved RFI classification sensitivity.

\begin{equation}
\frac{C}{N_{0}} > \frac{C}{N_{0} + I}
\label{eq:cn0_jam}
\end{equation}

The primary limitation of using C/N\textsubscript{0} as a stand-alone metric is its susceptibility to false detections. Physical obstructions, scintillation and multipath effects can produce signatures similar to those of jamming, potentially misleading detection algorithms. In addition, a spoofing system may adjust the power of spoofed signals and introduce noise to maintain nominal C/N\textsubscript{0} values. Consequently, C/N\textsubscript{0} must be considered alongside a received power metric to improve detection reliability and enable discrimination between different interference types.

\subsection{Spectrum Adjusted Received Power}
The received power metric is derived from the u-blox SPAN (signal characteristics) message, which  provides a Fast Fourier Transform (FFT) of the received signal with a frequency resolution of 500 kHz. The SPAN message output reflects the signal after it passes through the front-end RF chain and the receiver’s internal processing, requiring several processing steps to convert it into a usable metric. In addition, to calculate a signal’s spectral adjusted effective noise, the received power from each 500 kHz sub-band must be combined into a single value weighted by the expected GNSS signal’s Power Spectral Density (PSD). This conversion from a raw SPAN message to a useful metric involves four key processing steps: automatic gain control (AGC) adjustment, temperature calibration, single-value derivation, and conversion from the u-blox SPAN units to dBW/Hz units.

\subsubsection{AGC Adjustment}
The first step in the processing chain is adjusting the SPAN FFT data to account for the gain introduced by the Automatic Gain Control (AGC). The SPAN PGA message provides the AGC level, which is used to offset the applied gain and recover the pre-AGC power density. However, the information provided by receiver manufacturers is not always sufficient to perform this adjustment, as one AGC step does not correspond to a 1 dB change in the SPAN output. To address this limitation, u-blox F9P SPAN message outputs were analyzed during a GPS L2-adjacent RFI event at the GLONASS L2 center frequency, which changes the AGC level without adding noise to the GPS L2 frequency. By comparing the SPAN values before and after the AGC gain adjustment, a 3.7 dB offset was observed for each AGC step. To account for this offset, 3.7 times the PGA value (the u-blox variable for AGC step) is subtracted from the SPAN output, as shown in Equation~\ref{eq:agc_adj}.

\begin{equation}
SPAN_{adjusted} = SPAN - 3.7 \cdot PGA
\label{eq:agc_adj}
\end{equation}

\begin{figure}[htb]
\centering
\includegraphics[scale=0.28]{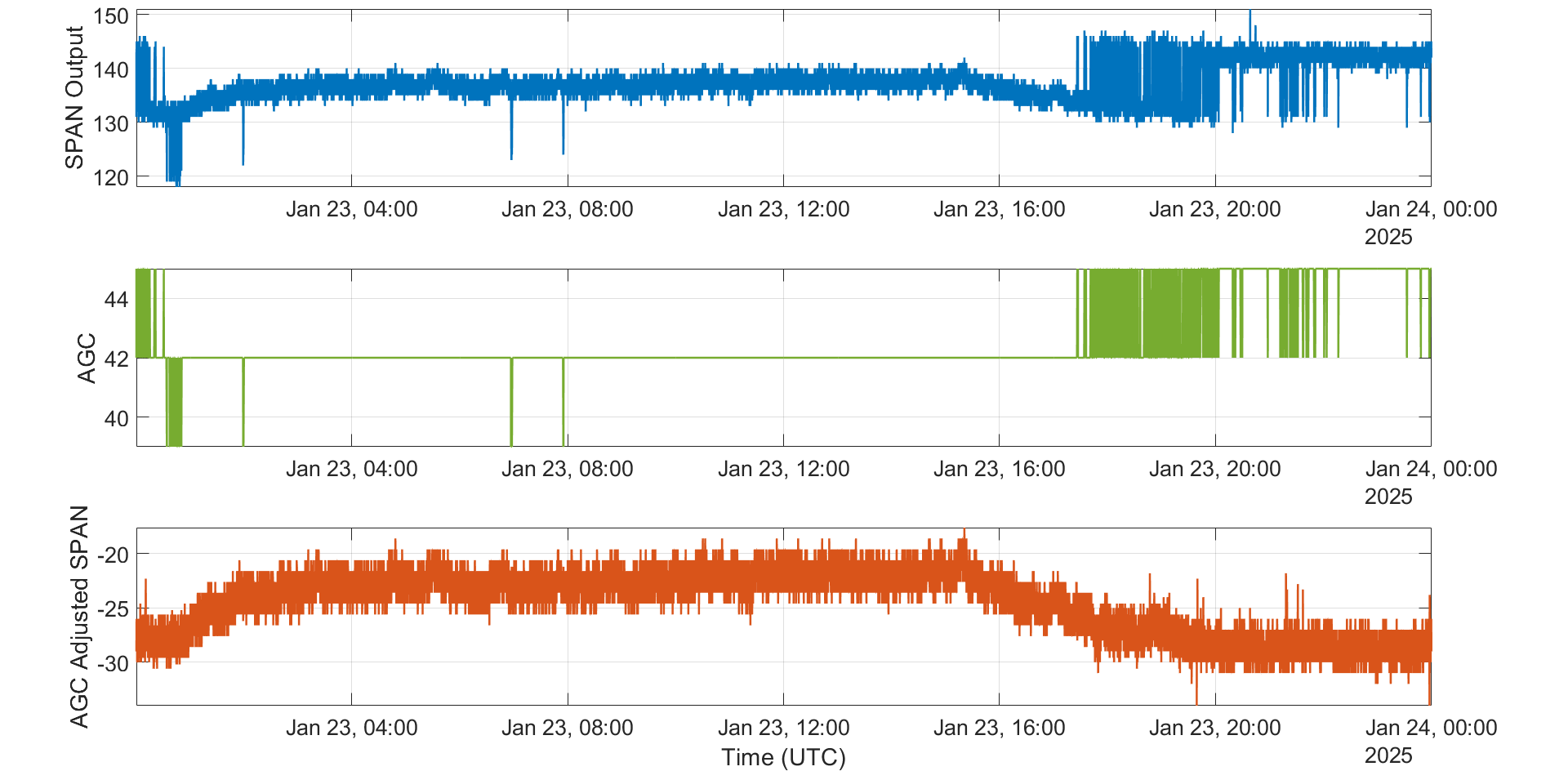}
\caption{GPS L2 SPAN message output for a single 500 kHz frequency bin centered at 1,227.5 MHz before AGC adjustment (blue), L2 band AGC value (green) and post-AGC adjusted SPAN output (orange)}
\label{fig:NAV25_AGC}
\end{figure}

Figure~\ref{fig:NAV25_AGC} shows the SPAN message output for a single 500 kHz frequency bin centered at 1,227.5 MHz (GPS L2), the L2-band AGC value, and the resulting post-adjustment output. The adjusted value accounts for the AGC-added gain, resulting in an output free of large step changes in power. It is worth noting that even after the AGC adjustment, the received power still fluctuates over a 24-hour period, this phenomenon is analyzed in the following sections.

\subsubsection{Temperature Calibration}
The second processing step involves calibration for external temperatures. While receivers installed indoors in temperature-controlled environments exhibit stable received power measurements, outdoor installations can experience large fluctuations. As seen in Figure ~\ref{fig:NAV25_AGC}, the AGC-adjusted received power values from the SPAN message over a 24-hour period for a receiver installed outdoors on Stanford's campus fluctuates. The plot, shown in UTC time, reveals a clear trend: during the daytime, when ambient temperatures are higher, the measured power is lower, conversely at night when temperatures drop the measured power increases.

\begin{figure}[htb]
\centering
\includegraphics[scale=0.29]{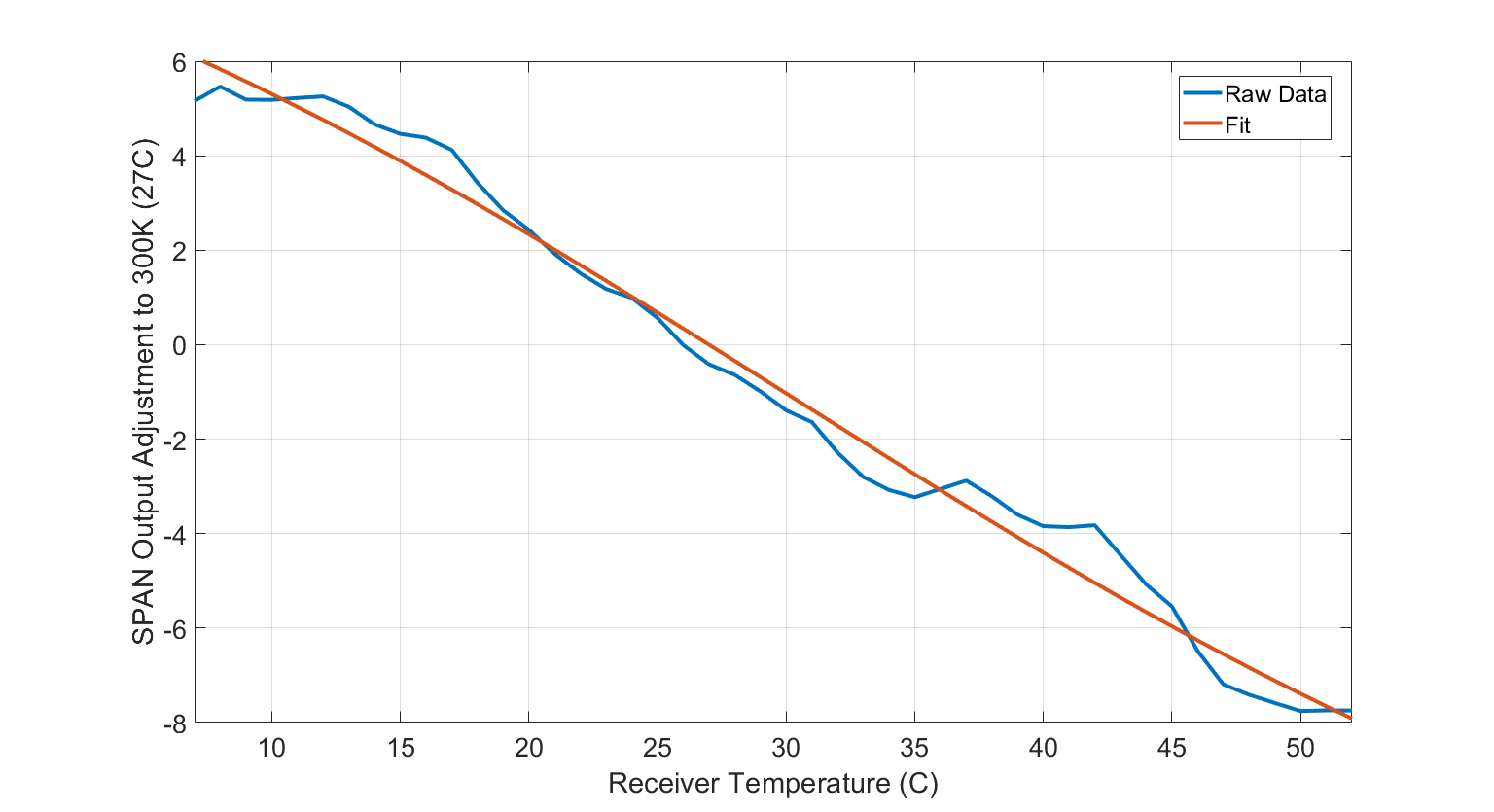}
\caption{Temperature calibration curve based on data sampled throughout the year, covering a temperature range of 7–52 C}
\label{fig:NAV25_temp_cal_curve}
\end{figure}

The changes in the receiver outputs can be attributed to the increased temperature degrading the performances of the receiver's amplifiers. To account for this effect, calibration was performed using data collected from outdoors receivers installed at Stanford, CA. The data were sampled from all seasons of the year to account for seasonal temperature variations, representing temperature cycles ranging from 7 to 52 degrees Celsius, and these values reflect the temperature at the receiver hardware as recorded by their internal sensors and not the ambient temperature. Figure ~\ref{fig:NAV25_temp_cal_curve} shows the calibration curve used to adjust the data points to a reference temperature of 300 Kelvin (27 Celsius) as well as the collected data. Over 300,000 raw data points, averaged for each temperature were used for the calibration curve fit.    

\begin{figure}[htb]
\centering
\includegraphics[scale=0.28]{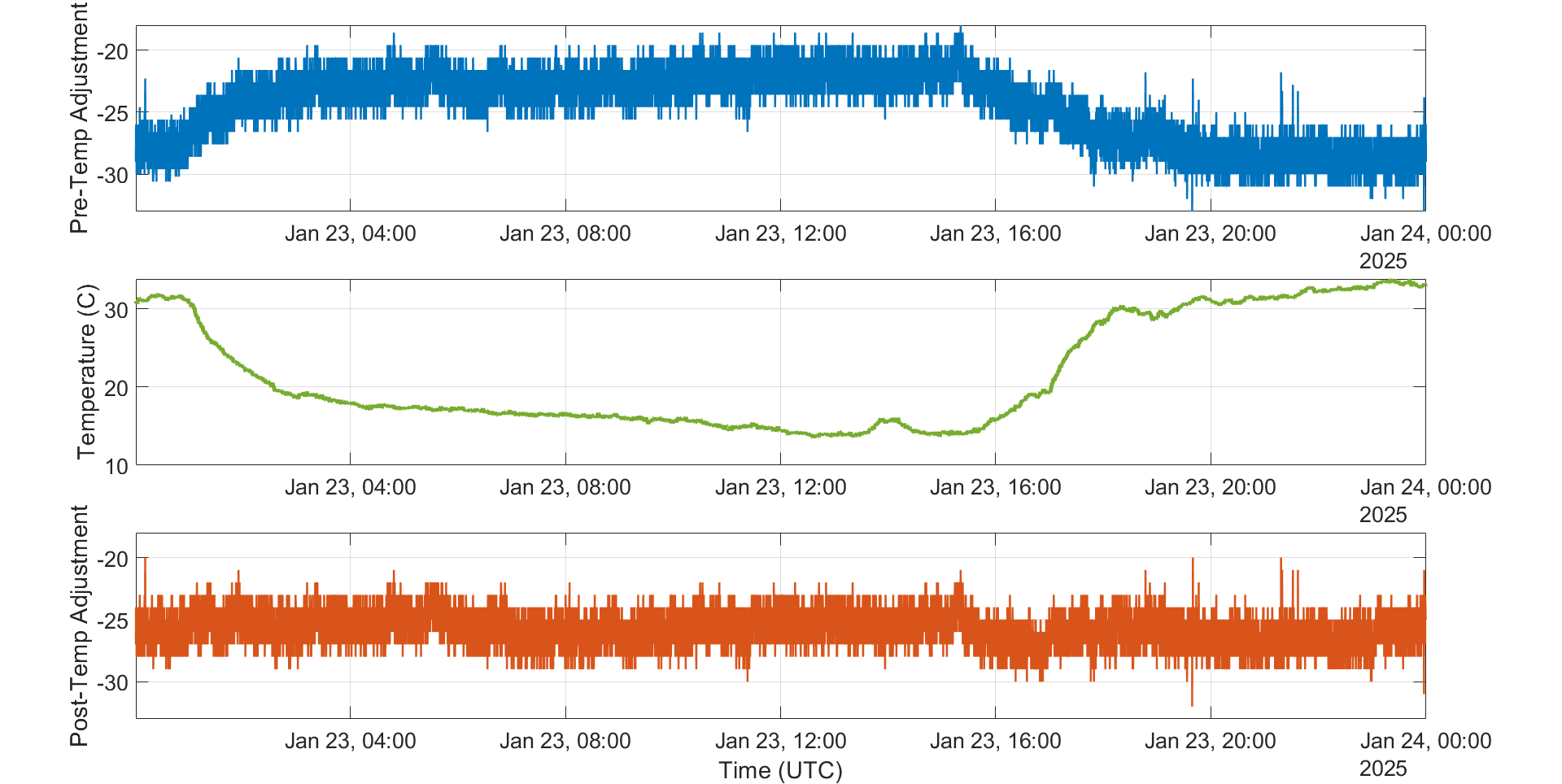}
\caption{GPS L2 AGC-adjusted SPAN output before temperature calibration (blue), receiver recorded temperature (green) and GPS L2 AGC-adjusted SPAN output after temperature calibration (orange)}
\label{fig:NAV25_temp_cal}
\end{figure}

Figure~\ref{fig:NAV25_temp_cal} presents the post-AGC-adjusted SPAN received power output before and after calibration to the reference temperature, alongside the receiver-recorded temperature value. While some fluctuation remains in the data, the calibration significantly reduces the temperature effect.

\subsubsection{Derive a Single Value for Rx Power}
The third step in processing the SPAN data involves deriving a single received power value from the set of 500 kHz FFT bins. GNSS signals vary in bandwidth and modulation schemes, for example, the GPS L1 C/A signal employs BPSK modulation centered at 1575.42 MHz. To ensure that the resulting received power metric is directly comparable to the signal’s C/N\textsubscript{0} value, the power from each 500 kHz bin weighted according to the Power Spectral Density (PSD) of the L1 C/A signal. This is accomplished by integrating the normalized received power density over the FFT bins, using the signal’s normalized PSD as a weighting function, as shown in Equation ~\ref{eq:ssc} \citep{RTCA2008DO235B} \citep{Kim2019GPSInterference}.

\begin{equation}
\kappa_{\text{S}, j} = \int_{-\beta/2}^{\beta/2} G_j(f) G_{\text{S}}(f) \, df
\label{eq:ssc}
\end{equation}

where:
\begin{itemize}
    \item $\kappa_{\text{S}, j}$: Spectral Separation Coefficient (SSC) between signal S and interference $j$ 
    \item $\beta_r \equiv$ Receiver bandwidth in Hz.
    \item $G_j(f) \equiv$ Unit-power spectra of interference $j$ in 1/Hz
    \item $G_{\text{S}}(f) \equiv$ Unit-power spectra of signal $S$ in 1/Hz
\end{itemize}

\begin{figure}[htb]
\centering
\includegraphics[scale=0.23]{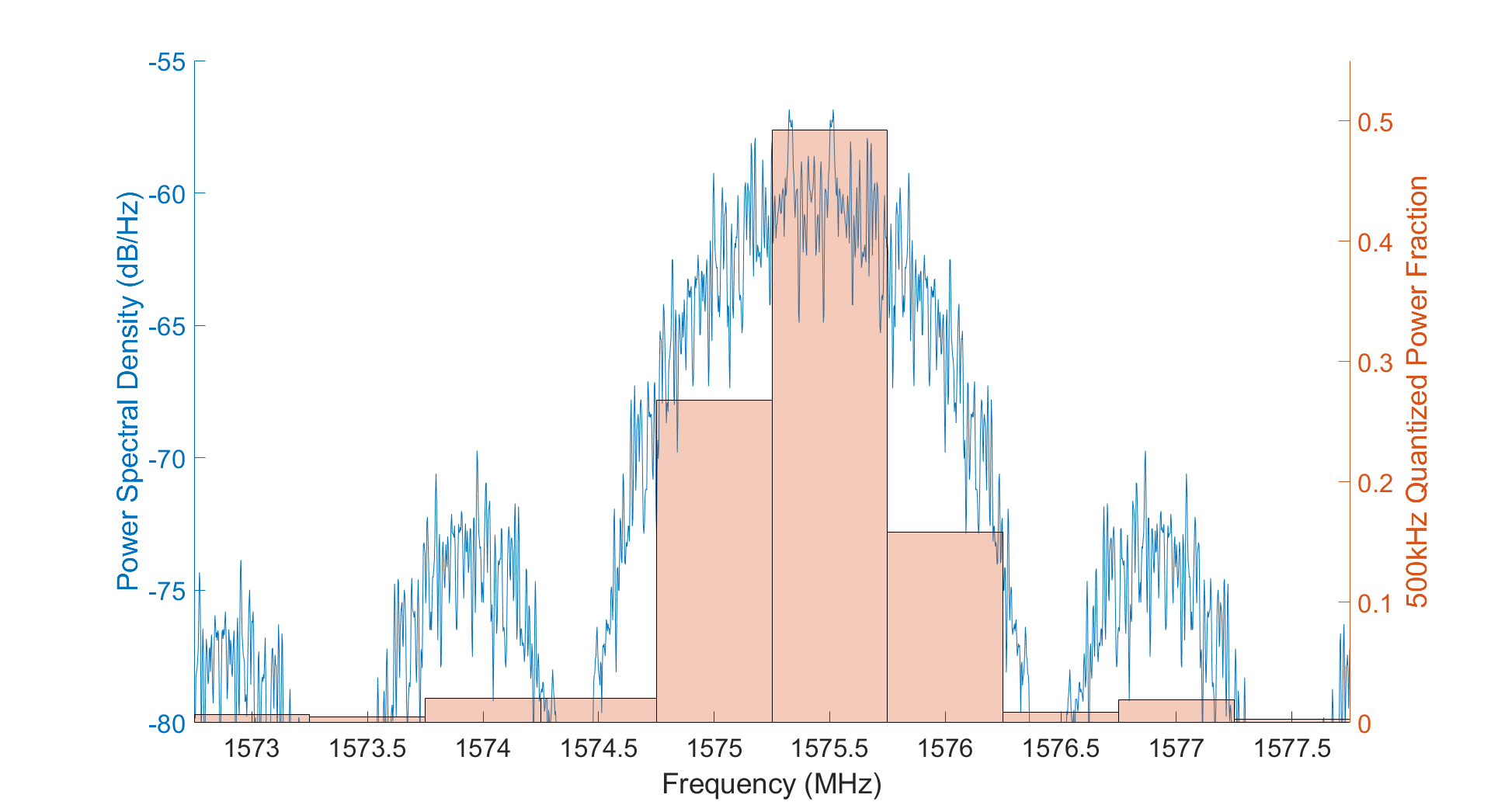}
\caption{GPS L1 C/A power spectral density overlaid by the quantized power fraction for a 500 kHz bin}
\label{fig:NAV25_psd}
\end{figure}

Since the signal’s PSD remains constant over time, the integration process can be simplified by precomputing a set of weighting factors for each 500 kHz bin within the SPAN message. These weights represent the relative contribution of each frequency bin to the total received power, based on the SSC of the bins and fixed PSD of the L1 C/A signal. Figure ~\ref{fig:NAV25_psd} presents the PSD of GPS L1 C/A overlaid by the quantized (500 kHz) power fraction. Weight factors can also be derived for other signals such as Galileo E1/E5 and GPS L5. Table ~\ref{tab:power_frac} presents the weighting factors used in the Rx Power calculation for each 0.5 MHz bin in the frequency range from 1573.0 MHz to 1577.5 MHz. The Doppler shift associated with the motion of the satellites can affect the weights of each bin, however, the change is negligible for the 0.5 MHz resolution of the u-blox F9P.

\begin{table}[htb]
 \caption{GPS L1 C/A Power Fraction per 0.5MHz}
 \label{tab:power_frac}
\begin{tblr}{colspec={X[l]X[c]X[c]X[c]X[c]X[c]X[c]X[c]X[c]X[c]X[c]},
width=\textwidth,
row{even} = {white,font=\small},
row{odd} = {bg=black!10,font=\small},
row{1} = {bg=black!20,font=\bfseries\small},
hline{Z} = {1pt,solid,black!60},
rowsep=3pt
}
Center Freq & 1573.0 & 1573.5 & 1574.0 & 1574.5 & 1575.0 & 1575.5 & 1576.0 & 1576.5 & 1577.0 & 1577.5\\
Power Fraction & 0.007 & 0.005 & 0.020 & 0.020 & 0.268 & 0.492 & 0.158 & 0.009 & 0.019 & 0.003 \\
\end{tblr}
\end{table}

\begin{figure}[htb]
\centering
\includegraphics[scale=0.28]{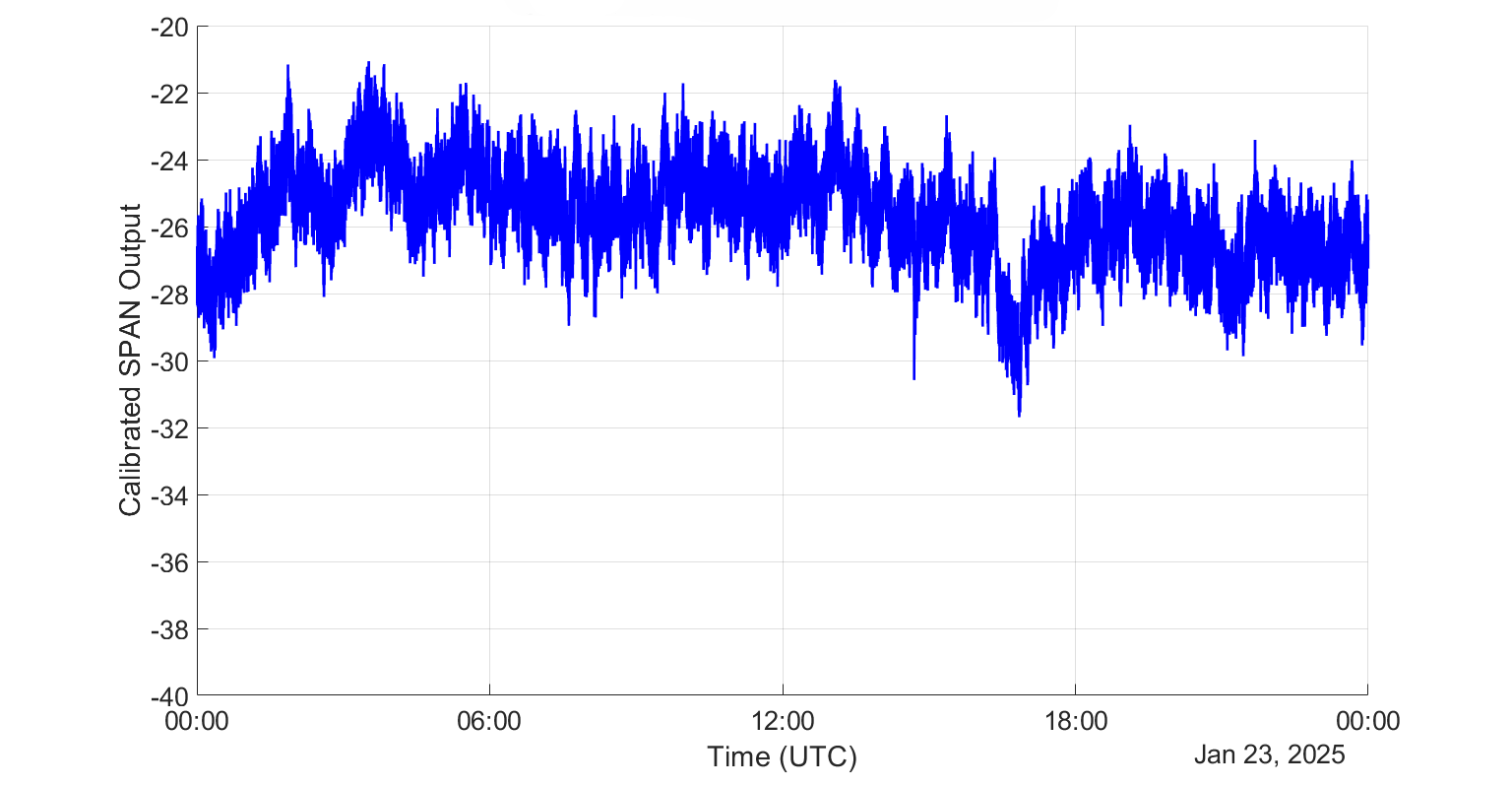}
\caption{Single GPS L1 C/A weighted values obtained by combining the 500 kHz bin SPAN outputs}
\label{fig:NAV25_single_power}
\end{figure}

Equation ~\ref{eq:add_power_frac} is used to convert the ten 500 kHz bin values, expressed in dB-scale, spanning from 1573.0 MHz to 1577.5 MHz, into a single received power estimate for the GPS L1 C/A signal. This is achieved by applying the predefined weighting factors derived from the signal’s Power Spectral Density. Figure ~\ref{fig:NAV25_single_power} shows the resulting GPS L1 C/A weighted received power obtained by summing the individual weighted 500 kHz bin values. 

\begin{equation}
\text{Rx power}_{\text{GPS L1 C/A}} = 10 \cdot log_{10}\sum_{i=1}^{n_b} 10^{P_i/10} f_i 
\label{eq:add_power_frac}
\end{equation}

where:
\begin{itemize}
    \item $n_b \equiv$ Number of 500kHz bins.
    \item $P_i \equiv$ Power of ith bin in dB/500kHz
    \item $f_i \equiv$ Power fraction of ith bin.
\end{itemize}

\subsubsection{Conversion to dBW/Hz unit}
\begin{figure}[htb]
\centering
\includegraphics[scale=0.52]{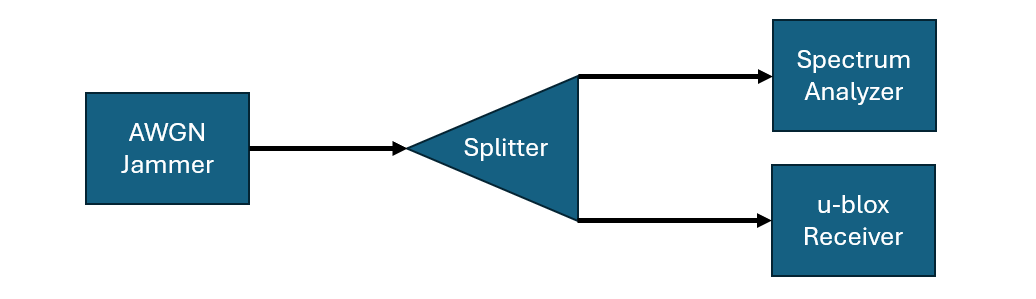}
\caption{Experiment setup for SPAN output to received power density conversion}
\label{fig:NAV25_lab_setup}
\end{figure}

The final step in the process involves converting the processed single-value SPAN output into units of dBW/Hz. Receiver manufacturers typically do not provide a direct method for translating FFT outputs into power density measurements suitable for assessing interference environments. To address this, a laboratory experiment was conducted to empirically establish the relationship between the SPAN outputs and actual power density levels. This calibration enabled the mapping of receiver-reported values to absolute power density units, allowing for meaningful interpretation of interference strength. As presented in Figure ~\ref{fig:NAV25_lab_setup}, the lab setup consisted of a noise source emitting wide-band Additive White Gaussian Noise (AWGN) to a u-blox F9P receiver. The noise generator feed was fed into both the receiver and a spectrum analyzer that can monitor the true power density. 

\begin{figure}[htb]
\centering
\includegraphics[scale=0.28]{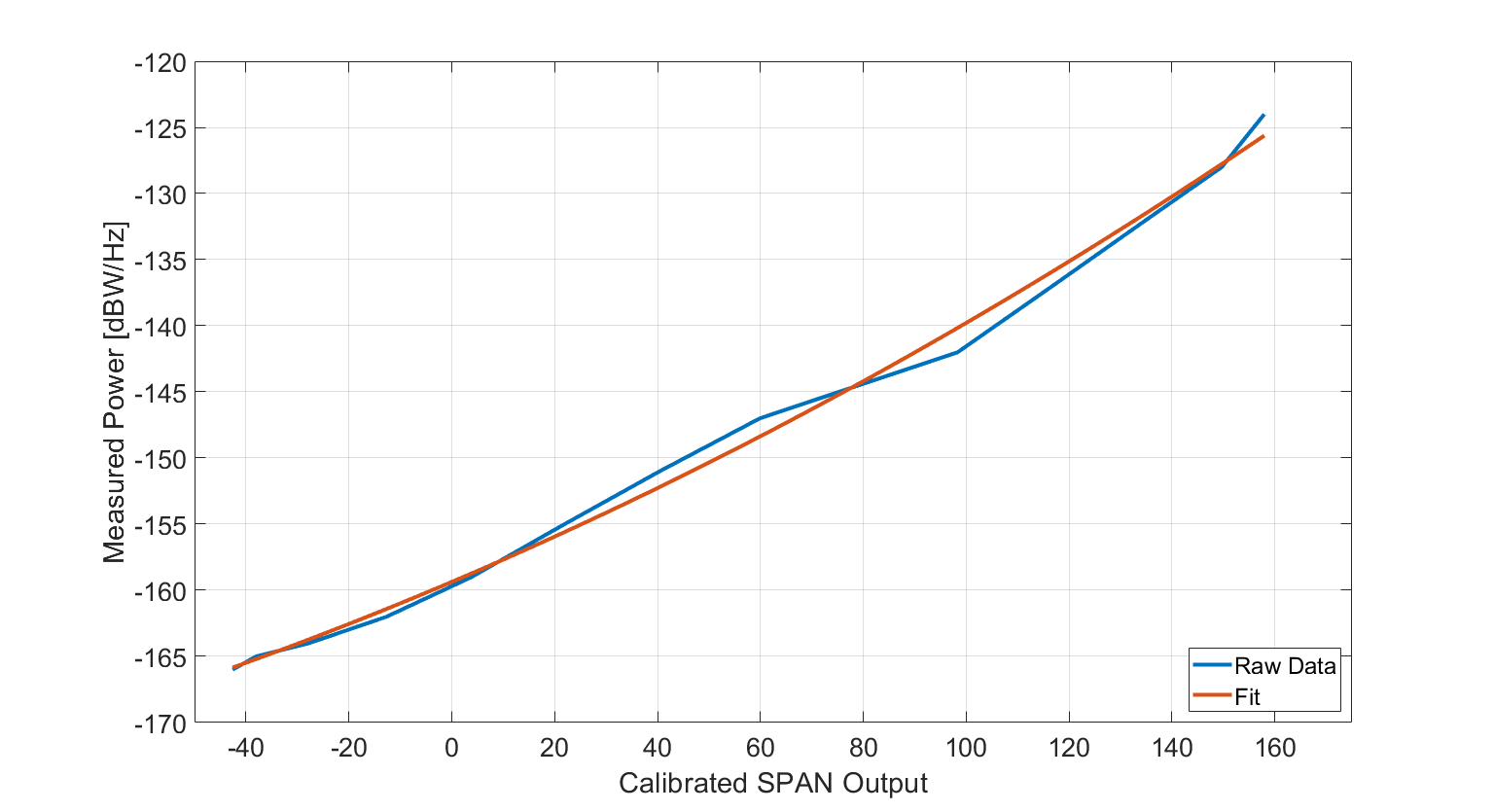}
\caption{Calibration curve of measured power density versus processed SPAN output}
\label{fig:NAV25_meas_power}
\end{figure}

The u-blox receiver outputs were recorded over time while the noise level was incrementally increased. These data were then used to derive a calibration curve that maps the processed SPAN output to the corresponding measured power density. Figure ~\ref{fig:NAV25_meas_power} presents the raw data collected during the experiment, along with the resulting curve fit. 

\begin{figure}[htb]
\centering
\includegraphics[scale=0.28]{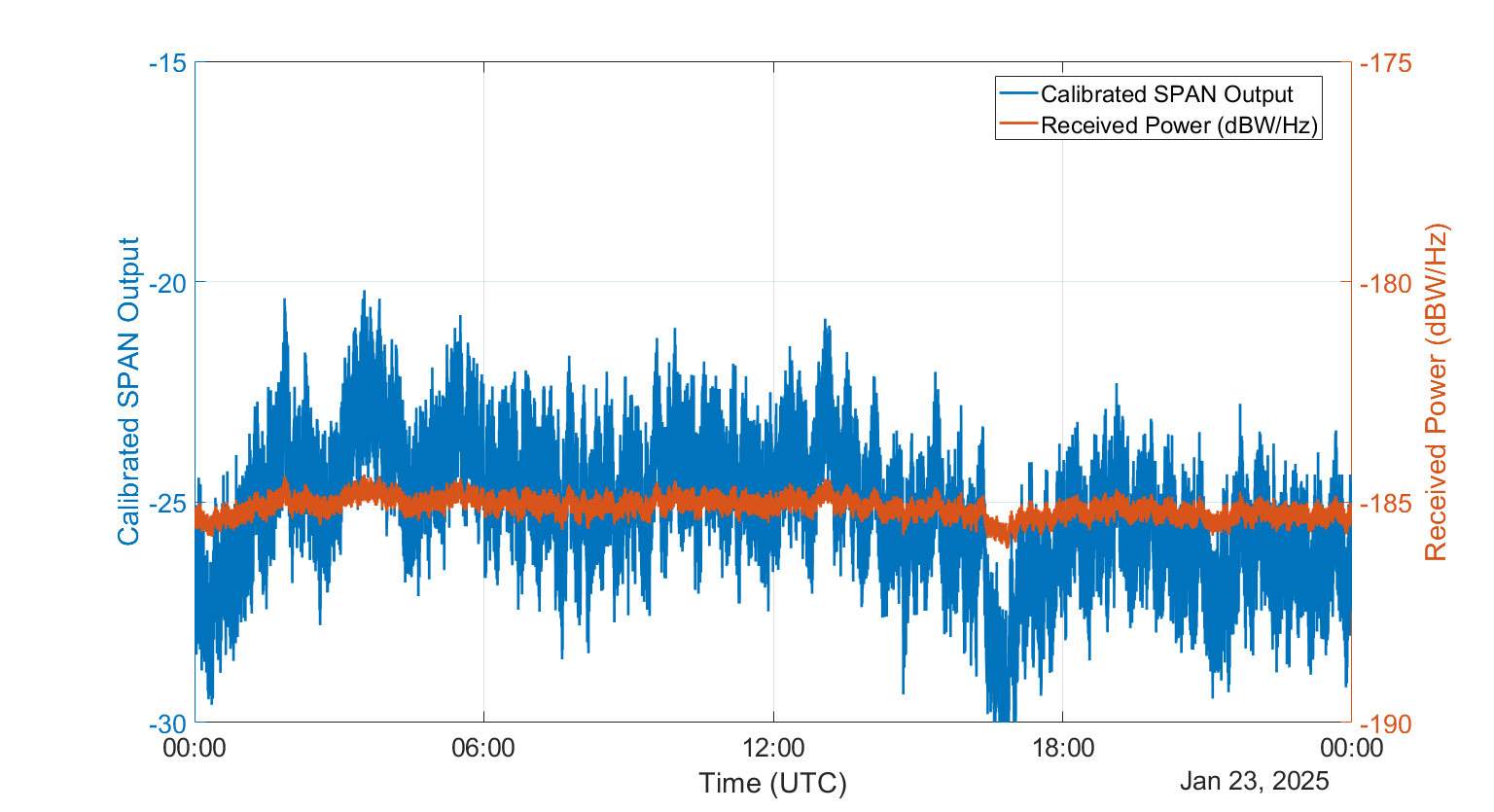}
\caption{GPS L1 C/A calibrated SPAN output (blue) and received power metric (orange) from RFI-free measurements collected at Stanford, CA}
\label{fig:NAV25_rx_power}
\end{figure}

The resulting number from the conversion to dBW-Hz units is for the power at the receiver input port. To calculate the received power at a given environment we need to account for the antenna and LNA gain as well as the cable loss between the antenna and the receiver. According to the specification sheet of the u-blox ANN-MB antenna, the total gain after accounting for the cable loss is 24.9 dB. As a result, the last part of the 4th step is to subtract the 24.9 dB gain. Figure ~\ref{fig:NAV25_rx_power} presents the received power calibrated for the GPS L1 C/A PSD from a receiver recording data on Stanford's campus. Since the GNSS signal power is below the noise floor, the received power remains relatively constant throughout the day, regardless of the number of satellites in-view. 

\subsection{C/N\textsubscript{0} over Received Power}
\begin{figure}[htb]
\centering
\includegraphics[scale=0.23]{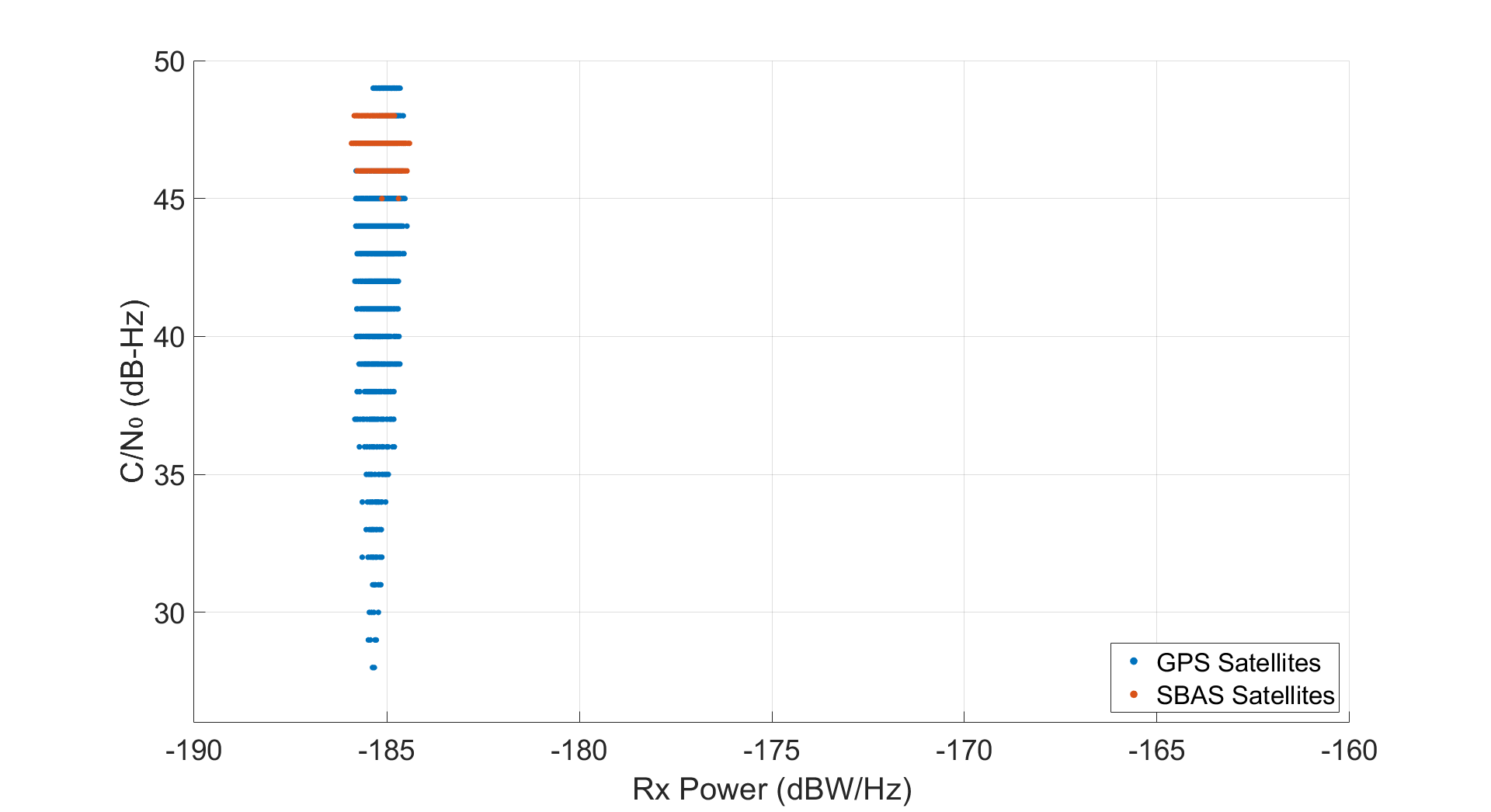}
\caption{C/N\textsubscript{0} over received power density distribution over a 24-hour observation at Stanford, CA}
\label{fig:NAV25_GPS_density}
\end{figure}

The combination of the C/N\textsubscript{0} and received power metrics provides insight not only into the presence of a disturbance, as indicated by deviations from nominal conditions, but also into its type. Figure~\ref{fig:NAV25_GPS_density} presents the C/N\textsubscript{0}-over-received-power distributions for GPS L1 C/A and SBAS L1 C/A signals over a 24-hour period under nominal conditions at Stanford University's campus. To enable a direct comparison, only measurements from GPS satellites at an elevation angle of 46° are included, matching the elevation angle of the SBAS satellite.

Given the tight cluster of SBAS C/N\textsubscript{0} measurements and the spectral adjustment of the received power metric to the GPS L1 C/A signal, the C/N\textsubscript{0} value is expected to decrease linearly as jamming power increases. To validate this hypothesis, a second laboratory experiment was conducted to characterize the response of the SBAS-based C/N\textsubscript{0} over received power metric under different jamming levels.

\begin{figure}[htb]
\centering
\includegraphics[scale=0.23]{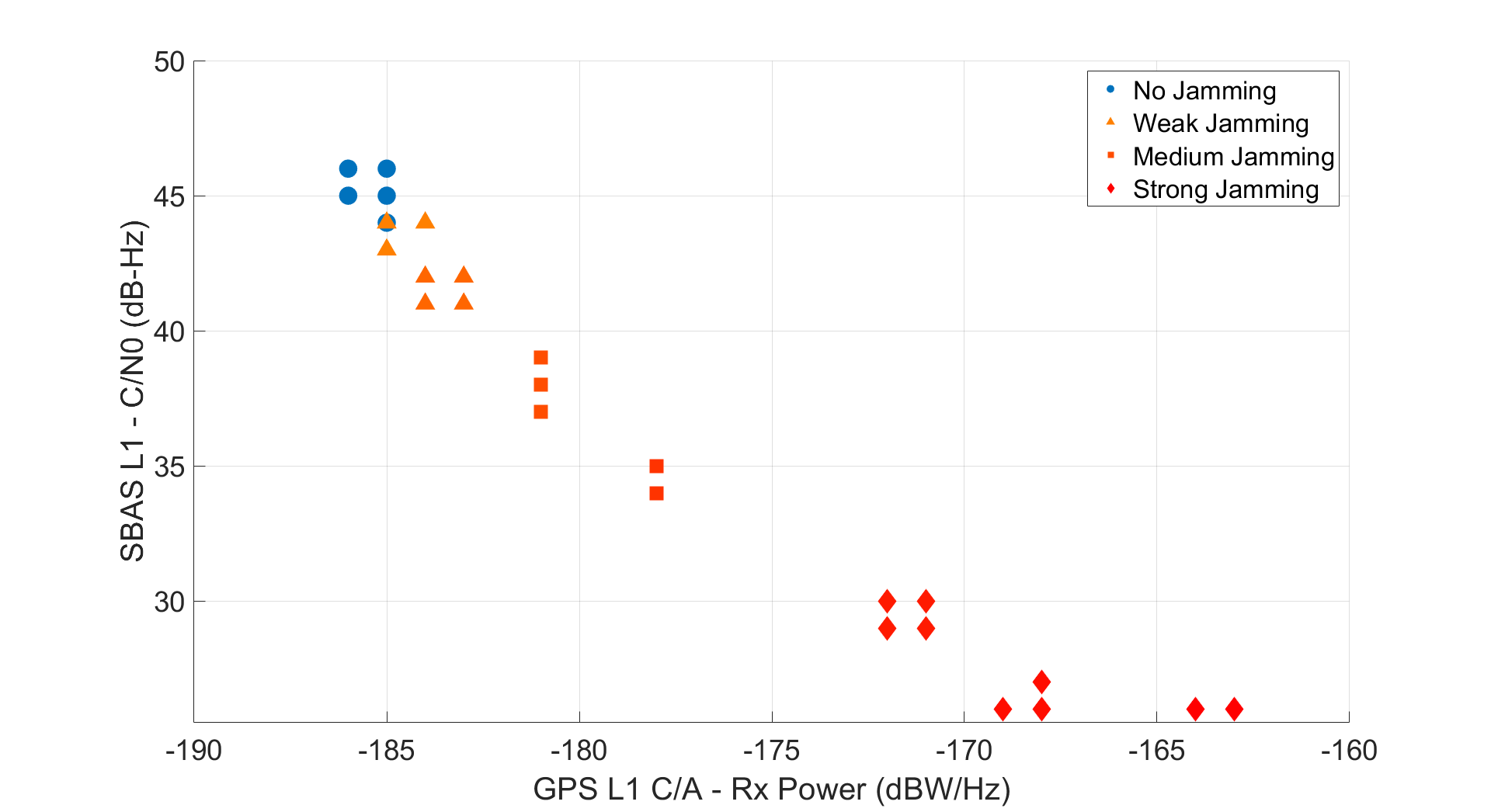}
\caption{C/N\textsubscript{0} over received power under lab-based increasing jamming conditions}
\label{fig:NAV25_SBAS_density_jamming}
\end{figure}

The experiment followed the setup described in Section 2.2.4, with the addition of a live-sky signal feed. Figure~\ref{fig:NAV25_SBAS_density_jamming} shows that the live-sky SBAS C/N\textsubscript{0} decreases by approximately 1 dB for every 1 dB increase in the GPS L1 C/A spectrum-adjusted received power. This linear relationship emerges only after the added noise exceeds the live-sky noise floor. Consequently, the metric is largely independent of the number of GPS satellites in view, since nominal GPS signals remain below the noise floor and do not dominate the measured input power. However, the received power metric is agnostic to the source of the additional noise. As a result, a new high-power RNSS signal introduced within the GPS L1 band could be indistinguishable from an RFI source.       

The linear degradation of C/N\textsubscript{0} with received power enables classification of disturbances into three categories: no RFI, jamming, and spoofing. Jamming follows the expected linear degradation trend, whereas spoofing yields elevated C/N\textsubscript{0} values relative to received power. The following section describes the methodology used to define the boundaries of each category in the C/N\textsubscript{0} over received power space. 

\section{RFI Threshold Definition}
Two thresholds need to be defined, one to detect an RFI disturbance and a second to classify the disturbance as either jamming or spoofing. Since the primary application of the proposed low-cost RFI monitoring system targets aviation environments, particularly airports, the thresholds are selected to achieve a low false-positive rate of one in one million (10\textsuperscript{-6}) per second, corresponding to approximately one false alarm every 12 days. Disturbance detection relies solely on the received power metric to avoid false detections caused by C/N\textsubscript{0} degradation from non-RFI effects such as scintillation or blockage. Disturbance characterization then exploits the relationship between C/N\textsubscript{0} and spectrum-adjusted received power, which are linearly related under jamming conditions.          

\subsection{RFI Detection}
\begin{figure}[htb]
\centering
\includegraphics[scale=0.225]{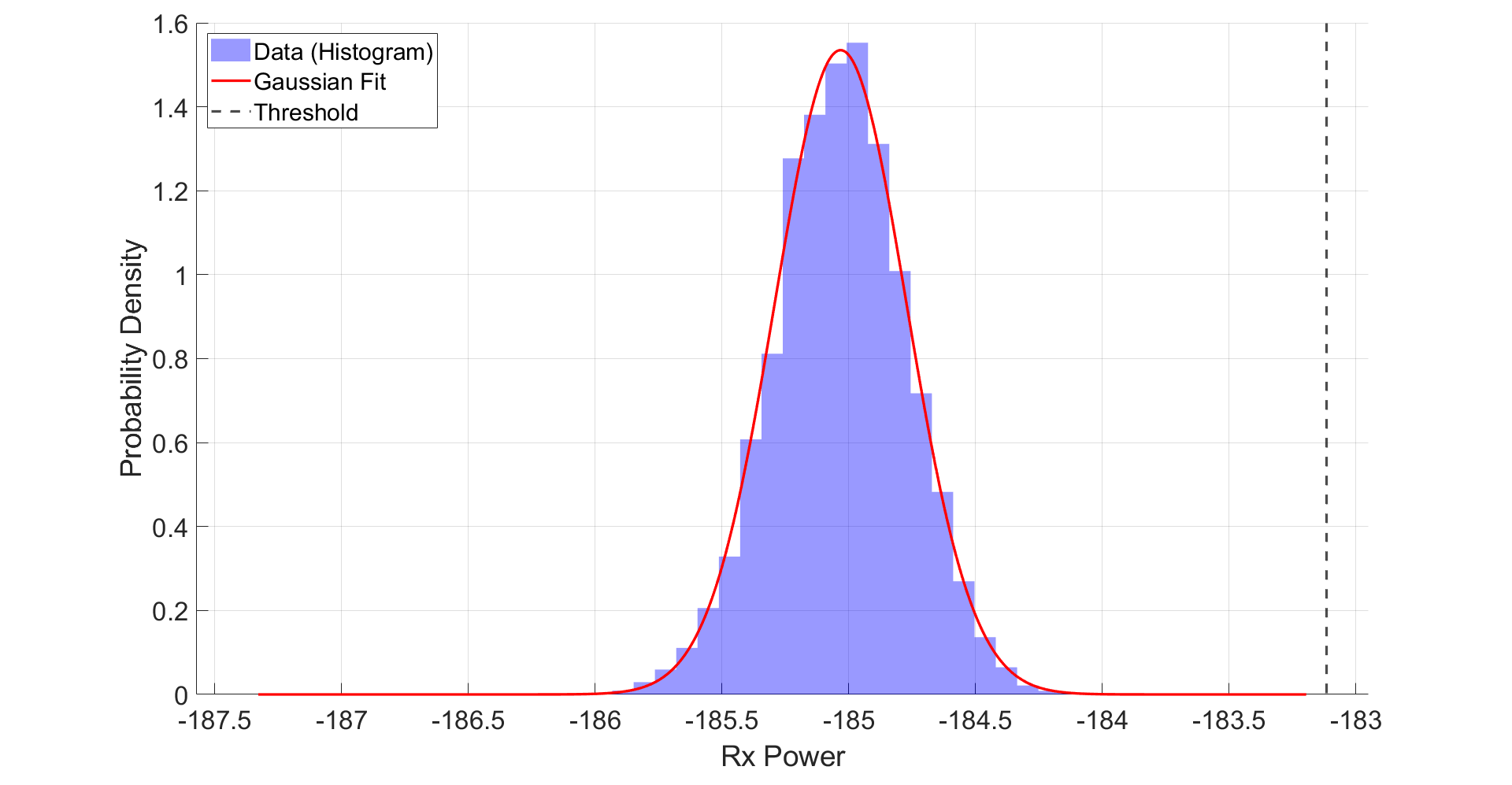}
\caption{Nominal received power data from 10 sample dates and the detection threshold}
\label{fig:NAV25_rx_power_threshold}
\end{figure}

To define the RFI detection threshold, data from 10 nominal days sampled throughout a year from receivers on Stanford’s campus were analyzed. The data were fitted with a Gaussian distribution, which was expanded to ensure all nominal data were overbounded. Using the standard deviation of the overbounded distribution and applying the 10\textsuperscript{-6} criterion, the detection threshold is defined as 1.92 dB from the nominal received power. Figure \ref{fig:NAV25_rx_power_threshold} presents the data distribution and the defined threshold.

Using the defined detection threshold, a test statistic is developed to evaluate the measured values and determine whether a disturbance exists and the extent to which it exceeds the threshold.   

\begin{equation}
T_{\mathrm{RFI}}(\mathbf{x}) = \frac{\mu_{rx_{power}} + \gamma_{rx_{power}}}{x},
\qquad
T_{\mathrm{RFI}}
\mathop{\gtrless}_{H_0}^{H_1}
1.
\end{equation}

where $x$ is the measured received power value in dB units, $\mu$ is the mean received power of a location and $\gamma_{rx_{power}}$ is the detection threshold of 1.92 dB.

\subsection{RFI Classification}
To classify an RFI disturbance as either jamming or spoofing, the relationship between a signal's C/N\textsubscript{0} and received power must be monitored. As shown in Figure \ref{fig:NAV25_SBAS_density_jamming}, C/N\textsubscript{0} decreases linearly as received power increases. To define the SBAS C/N\textsubscript{0} ceiling, the same methodology used for the received power threshold is applied, resulting in a threshold of 3 dB above the nominal median C/N\textsubscript{0} value.

\begin{figure}[htb]
\centering
\includegraphics[scale=0.22]{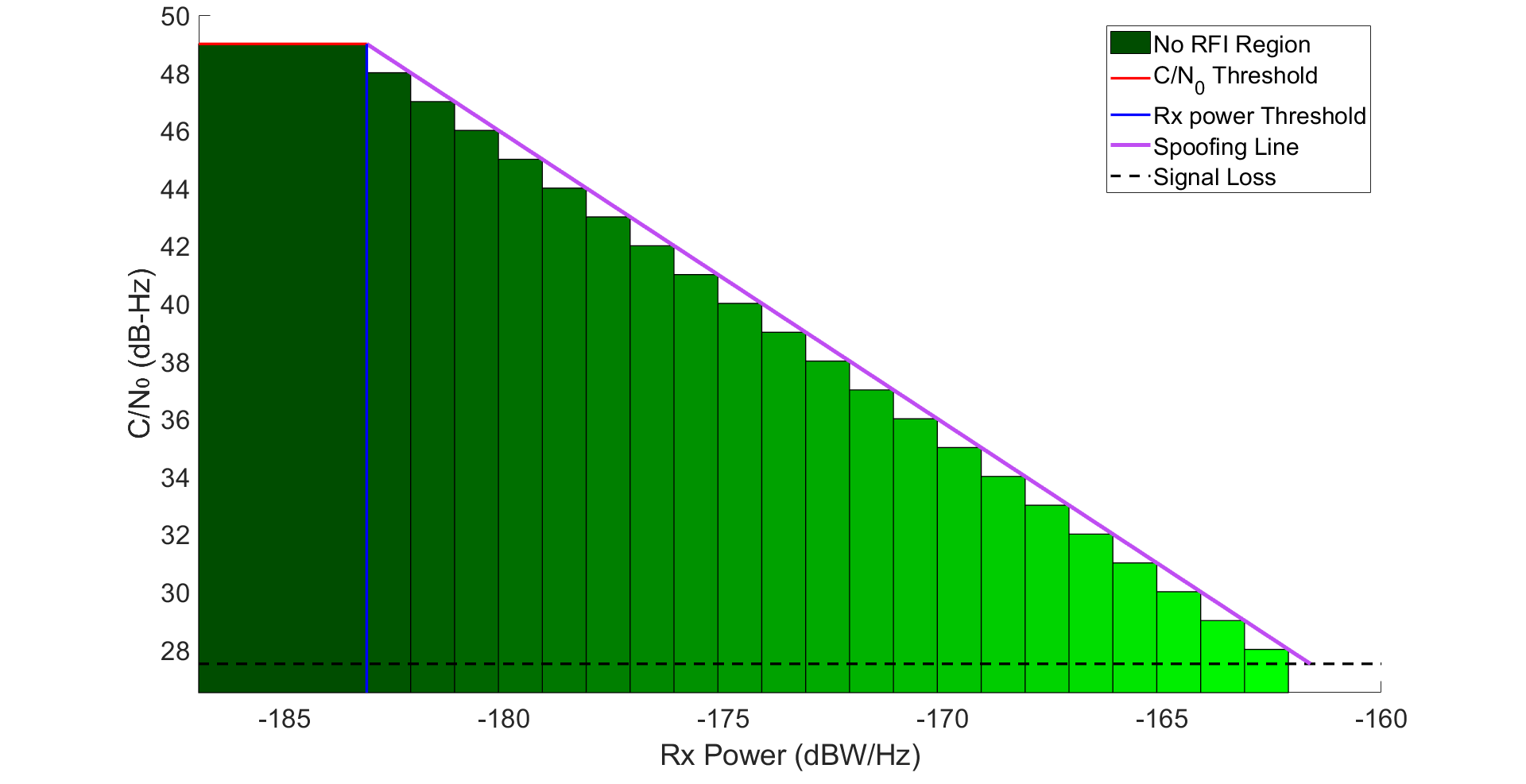}
\caption{Spoofing threshold definition for different received power values}
\label{fig:NAV25_spoof_threshold}
\end{figure}

Since the spoofing detection threshold depends on received power, it must decrease as received power increases. Figure \ref{fig:NAV25_spoof_threshold} presents the resulting spoofing detection threshold as a function of received power up to signal loss, which is defined as 27 dB-Hz for the purposes of this paper.

The spoofing threshold line can be defined as a test statistic, to implemented in the detection and classification algorithm. 

\begin{equation}
T_{\mathrm{spoofing}}(\mathbf{x}, \mathbf{y}) = \frac{(\mu_{rx_{power}} + \gamma_{rx_{power}}) + (m_{C/N_0} + \gamma_{C/N_0})}{x+y},
\qquad
T_{\mathrm{spoofing}}
\mathop{\gtrless}_{H_0}^{H_1}
1.
\end{equation}

where $x$ is the measured received power value in dB units, $\mu$ is the mean received power of a location, $y$ is the measured C/N\textsubscript{0} value, $m$ is the median C/N\textsubscript{0} value and $\gamma_{rx_{power}}$ / $\gamma_{C/N_0}$ are the detection threshold of 1.92 dB for recevied power and 3 dB for SBAS C/N\textsubscript{0}. 

\begin{figure}[htb]
\centering
\includegraphics[scale=0.22]{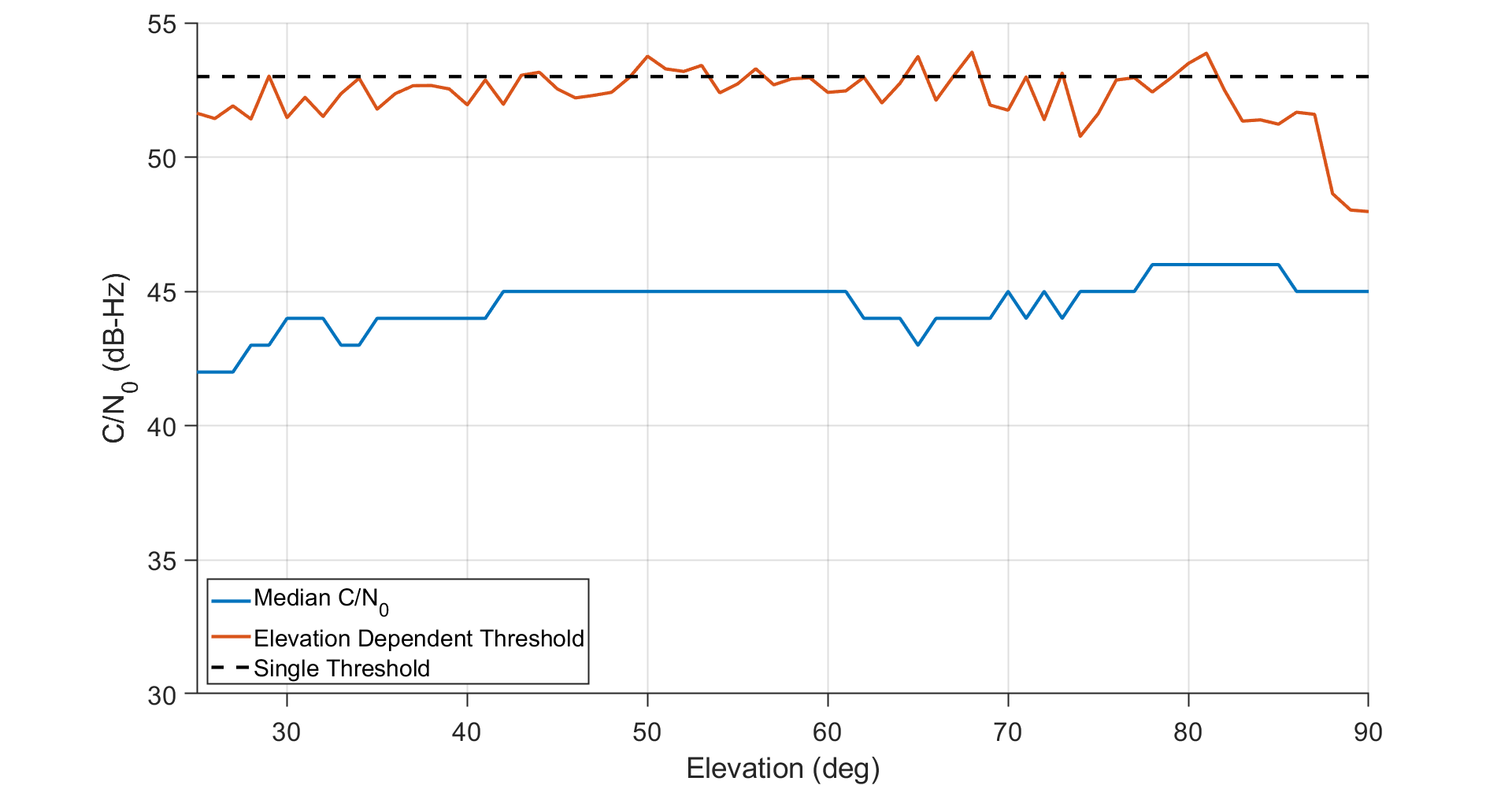}
\caption{Median GPS C/N\textsubscript{0} over elevation angle and corresponding threshold}
\label{fig:NAV25_GPS_threshold}
\end{figure}

While SBAS signals provide higher-fidelity spoofing classification due to their lower C/N\textsubscript{0} variability, GPS signals can still be used to classify spoofing on individual GPS satellites. For GPS signals, the C/N\textsubscript{0} distribution is asymmetric, with greater variability below the median than above it. This behavior is expected, as several effects, including blockage and multipath, can reduce signal strength, whereas the maximum C/N\textsubscript{0} is constrained by the transmitted signal power. To account for this asymmetry, the Gaussian fit and threshold definition are performed using only C/N\textsubscript{0} values above the median. A separate threshold is then defined for each elevation angle to account for the elevation-dependent propagation loss. Figure \ref{fig:NAV25_GPS_threshold} presents the median C/N\textsubscript{0} as a function of elevation angle and the corresponding thresholds. Since the C/N\textsubscript{0} threshold is nearly constant across all elevation angles, a single threshold can be used for GPS spoofing classification. Consequently, the same test statistic defined for SBAS signals can be applied, with $(\mu_{C/N_0} + \gamma_{C/N_0})$ set to 53 dB-Hz.    

\subsection{C/N\textsubscript{0} over Received Power Regions and Calibration}
\begin{figure}[htb]
    \centering

    \begin{subfigure}{\linewidth}
        \centering
        \includegraphics[scale=0.315]{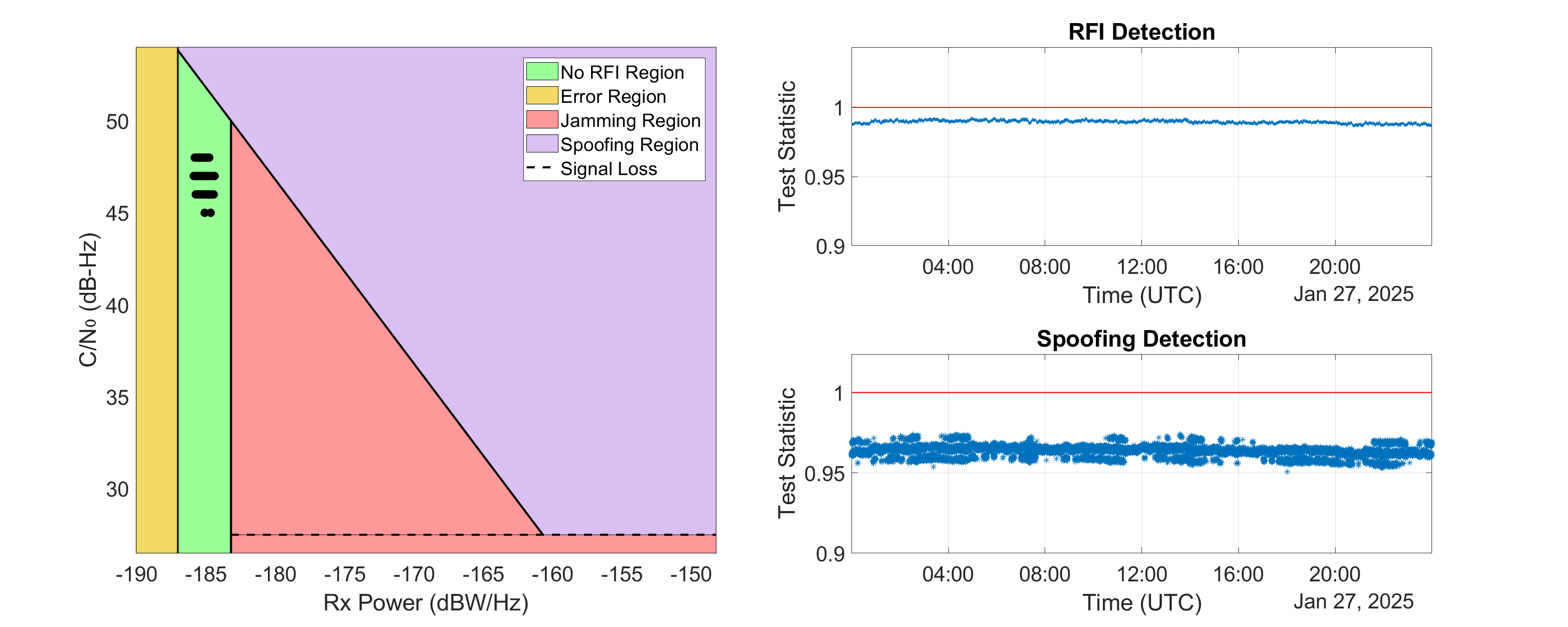}
        \caption{SBAS C/N\textsubscript{0} over received power metric}
        \label{fig:Nav25_nominal_sbas}
    \end{subfigure}

    \vspace{0.3em}

    \begin{subfigure}{\linewidth}
        \centering
        \includegraphics[scale=0.315]{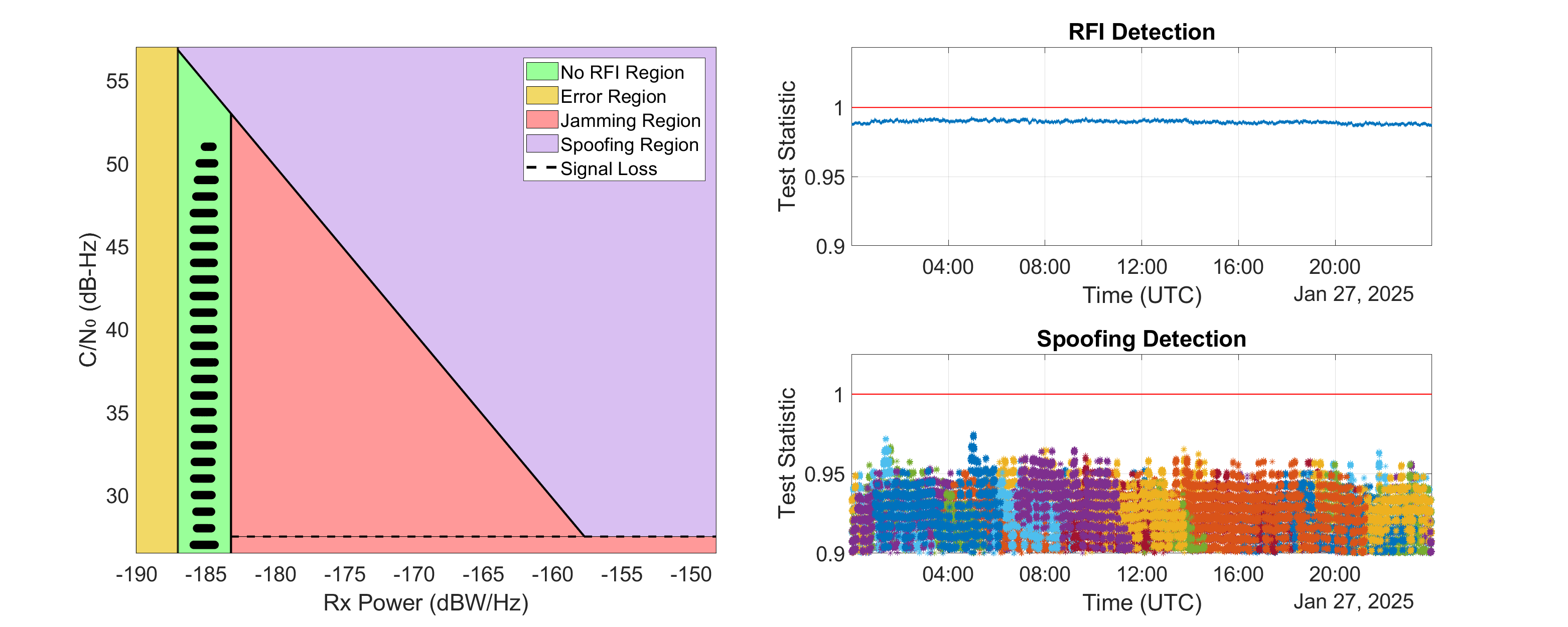}
        \caption{GPS C/N\textsubscript{0} over received power metric}
        \label{fig:Nav25_nominal_gps}
    \end{subfigure}

    \caption{C/N\textsubscript{0} over received power metric from 24 hours of nominal data}
    \label{fig:Nav25_nominal}
\end{figure}

The resulting C/N\textsubscript{0} over received power metric is two-dimensional and consists of three main regions: no RFI, jamming, and spoofing. The no-RFI region is bounded by the received power and maximum C/N\textsubscript{0} thresholds, where both test statistics are negative. The jamming region is bounded by the same thresholds and is characterized by a positive RFI test statistic and a negative spoofing test statistic. The spoofing region lies above the spoofing threshold, where the spoofing test statistic is positive. A fourth region is defined for received power values below the calibrated noise floor to detect anomalies in the receiver setup and trigger recalibration or maintenance. Figure \ref{fig:Nav25_nominal} presents the RFI detection and classification metric during 24 hours of nominal data from Stanford's campus. Figure \ref{fig:Nav25_nominal_sbas} includes the results based on an SBAS signal and \ref{fig:Nav25_nominal_gps} includes the results  from the available GPS signals. Since there are multiple GPS satellites in-view at any given time, the spoofing test statistic can be evaluated for each individual satellite.   

While the relative threshold boundaries remain unchanged regardless of receiver deployment location, their centering must be calibrated for each new installation. For the SBAS metric, both the nominal received power and the median SBAS C/N\textsubscript{0} must be estimated. For the GPS metric, the median C/N\textsubscript{0} at each available elevation angle is estimated and used to define the corresponding thresholds. Although longer calibration periods improve confidence, as little as 30 minutes of nominal data are sufficient to establish an initial baseline.

The availability of the received power metric enables the calibration process to be fully automated without manual user input. The algorithm first identifies nominal no-RFI samples by finding the minimum received power and selecting all samples within 3.84 dB of that value, corresponding to the width of the no-RFI region. The mean received power of the selected samples is then calculated and assigned as $\mu_{rx_{power}}$. Using the same nominal samples, the median SBAS C/N\textsubscript{0} and the median GPS C/N\textsubscript{0} at each available elevation angle are computed. The GPS C/N\textsubscript{0} thresholds are then obtained by comparing the median C/N\textsubscript{0} values at each elevation with the reference median values shown in Figure \ref{fig:NAV25_GPS_threshold}. All results presented in this paper use the automated calibration procedure.          

\section{Real World Validation}
The proposed RFI detection and classification methodology, developed using data collected at Stanford University, was validated using live-sky RFI data from the 2024 Norway Jammertest. Held annually each September in Bleik, Norway, the Jammertest provides a controlled yet realistic environment for evaluating RFI monitoring techniques \citep{JammerTest2026}. The interference sources and their characteristics are known, enabling a detailed assessment of the proposed methodology. Each test day consists of a series of jamming and spoofing scenarios with predefined transmit power levels. This study uses data collected on September 9 and 10, with the first day dedicated to jamming and the second to spoofing in the form of meaconing. The transmission scenarios from both days are summarized in Table \ref{tab:jamming_scenarios}.

\begin{table}[htb]
 \caption{Jammertest day 1 and day 2 RFI scenarios}
 \label{tab:jamming_scenarios}
\begin{tblr}{
colspec={X[c]X[c]X[c]X[c]X[c]X[c]},
width=\textwidth,
row{even} = {white,font=\small},
row{odd} = {bg=black!10,font=\small},
row{1} = {bg=black!20,font=\bfseries\small,valign=m},
hline{Z} = {1pt,solid,black!60},
vline{4} = {1.2pt, solid, black},
rowsep=3pt
}
Day 1 Time (UTC) & Day 1 Test & Details & Day 2 Time (UTC) & Day 2 Test & Details \\
\hline
10:00   & Recording start  & -     & 6:40          & Recording start   &  - \\
12:10-12:20 &  Jamming & 50W CW L1   & 7:00-7:05   & Spoofing (Meacon)        & 1W Meacon (RX1) \\
12:25-12:35 &  Jamming & 50W CW L1, G1 & 7:15-7:25   & Jamming and Spoofing (Meacon)  & 1W Meacon (RX1) with initial jamming  \\
12:40-12:50 &  Jamming & 50W Sweep L1 (6 MHz) & 7:35-7:40   & Spoofing (Meacon) & 10W Meacon (RX1) \\
13:00-13:10 &  Jamming & 50W Sweep L1, G1 (6 MHz) &  7:50-8:00   & Jamming and Spoofing (Meacon) & 10W Meacon (RX1) with initial jamming \\
13:20-13:30 &  Jamming & 50W PRN L1 & 8:10-8:15   & Spoofing (Meacon)         & 10W Meacon (RX2) \\
13:40-13:50 &  Jamming & 50W PRN L1, G1  & 8:25-8:45  & Spoofing (Meacon) and Jamming & 10W Meacon (RX1/RX2 switch) then  jamming \\
 14:00 & Recording End     & - & 8:55-9:05 & Spoofing (Meacon) & 10W Meacon (RX1/RX2 switch) \\
 &      &    & 9:08   & Recording End          & - \\
\end{tblr}
\end{table}

The same receiver and antenna configuration was used at both Stanford University and the Jammertest, consisting of a u-blox F9P receiver and an ANN-MB antenna. Since SBAS SV 131, used at Stanford, is not visible over Europe, EGNOS SV 136 was used instead. Furthermore, only live-sky and laboratory data collected at Stanford were used to develop the detection thresholds, therefore, the evaluation presented here is entirely independent of the development dataset. As described in the previous section, threshold calibration was performed autonomously using the proposed calibration algorithm.

\subsection{Jammertest Day 1 - Jamming}

\begin{figure}[htb]
    \centering

    \begin{subfigure}{\linewidth}
        \centering
        \includegraphics[scale=0.25]{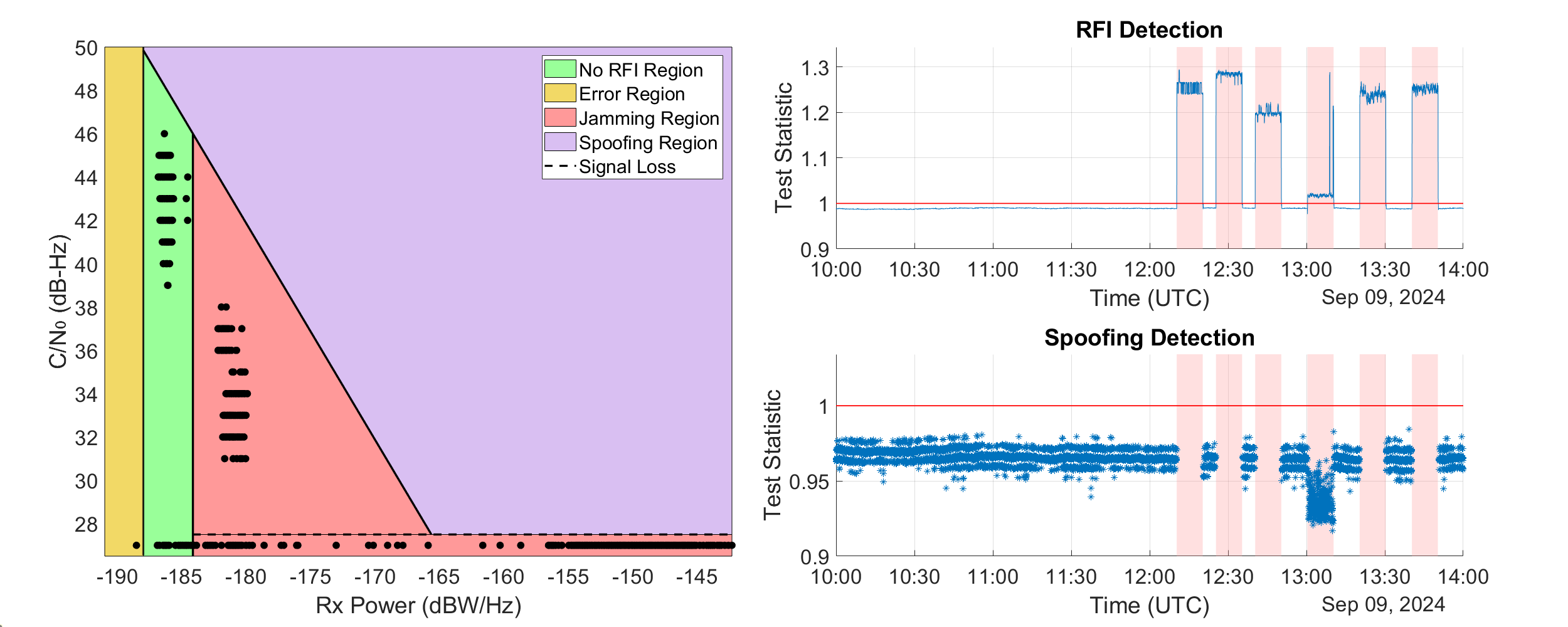}
        \caption{SBAS C/N\textsubscript{0} over received power metric}
        \label{fig:Nav25_jam1_sbas}
    \end{subfigure}

    \vspace{0.3em}

    \begin{subfigure}{\linewidth}
        \centering
        \includegraphics[scale=0.25]{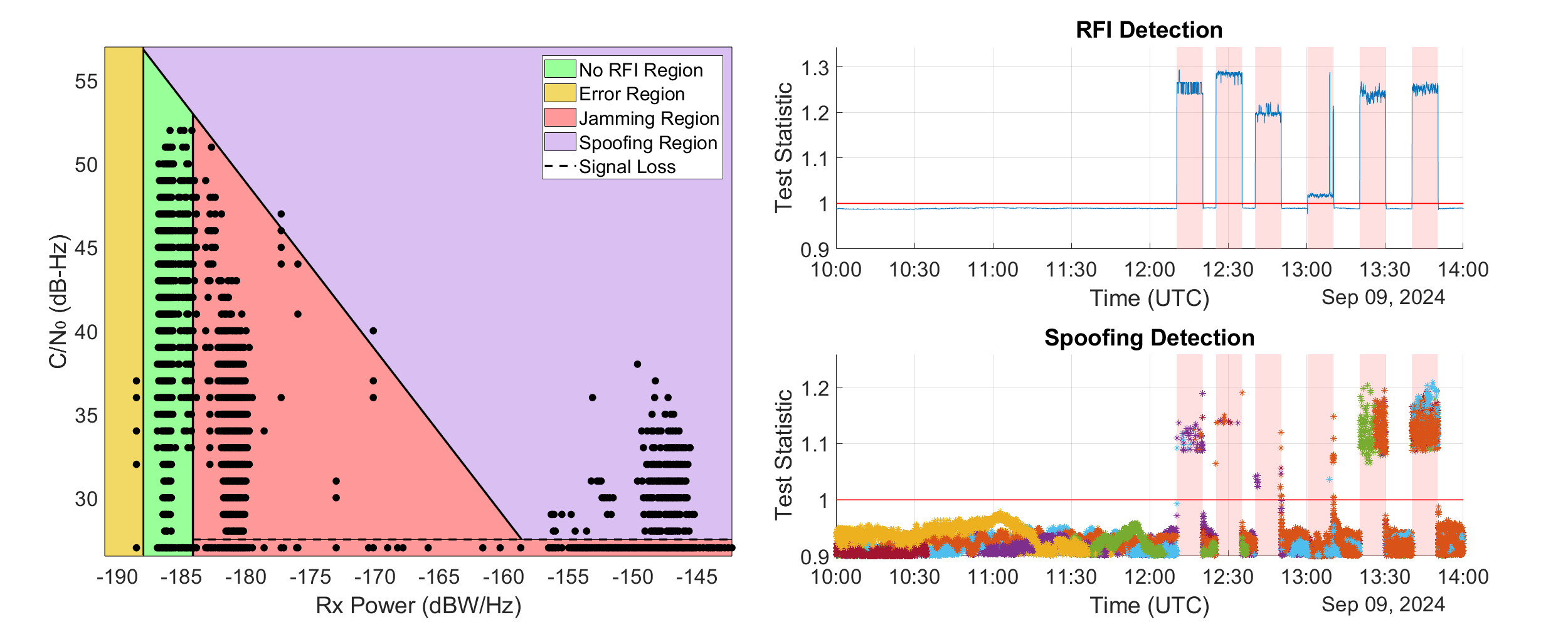}
        \caption{GPS C/N\textsubscript{0} over received power metric}
        \label{fig:Nav25_jam1_gps}
    \end{subfigure}

    \caption{C/N\textsubscript{0} over received power metric from Day 1 of Jammertest (Receiver 1)}
    \label{fig:Nav25_Jam_Day1}
\end{figure}

Figure~\ref{fig:Nav25_Jam_Day1} presents results from one of the four receivers deployed during the jamming test on 9 September, selected as a representative example. Results for the remaining receivers are summarized in Table~\ref{tab:jam_test_matrix}. The figure is based on a four-hour recording collected that day. The left panel shows the two-dimensional C/N\textsubscript{0}-over-received-power metric, while the right panel shows the RFI and spoofing test statistics as a function of time. Jamming periods, representing ground truth, are highlighted in red.

Detailed RFI flag results for each receiver are provided in Table~\ref{tab:jam_test_matrix}. The first column lists the RFI scenarios, grouped by RFI type. For each scenario, the table reports the percentage of time each receiver flagged that condition, with the ground-truth cell shaded green. Inspection of the "No RFI Flag" column shows that the algorithm correctly identified the presence of RFI with better than 99\% accuracy. The small number of false positives and false negatives is primarily attributable to receiver response time at the onset and termination of jamming events. These results demonstrate the effectiveness of the predefined thresholds even when deployed in a new environment.

\begin{table}[!htbp]
\caption{Jammertest Day 1 (Jamming) resulting RFI flags, with ground truth highlighted in green}
\label{tab:jam_test_matrix}
\begin{tblr}{
  colspec = {X[2.2,l] X[1,c] X[1,c] X[1,c] X[1,c] X[1,c] X[1,c]},
  width = \textwidth,
  row{1-2} = {bg=black!20, font=\bfseries\small, valign=m}, 
  row{3-6}   = {bg=black!10, font=\small}, 
  row{7-10}  = {white, font=\small},       
  row{11-14} = {bg=black!10, font=\small}, 
  row{15-18} = {white, font=\small},       
  cell{3-6}{3}   = {bg=green!15},
  cell{3-6}{4-7}   = {bg=orange!15},           
  cell{7-18}{4-5} = {bg=green!15},
  cell{7-18}{3}   = {bg=orange!15},
  cell{7-18}{6-7}   = {bg=orange!15},
  hline{Z} = {1pt, solid, black!60},
  rowsep = 0.75pt
}
\SetCell[r=2]{l} RFI Scenario & \SetCell[r=2]{c} Receiver & \SetCell[r=2]{c} No RFI Flag & \SetCell[c=2]{c} Jamming Flag & & \SetCell[c=2]{c} Spoofing Flag & \\
& & & SBAS & GPS & SBAS & GPS \\ 
\SetCell[r=4]{l} No Testing & Receiver 1 & 99.7\% & 0.3\% & 0.3\% & 0.0\% & 0.0\% \\
& Receiver 2 & 99.6\% & 0.3\% & 0.4\% & 0.1\% & 0.0\% \\
& Receiver 3* & 99.2\% & 0.8\% & 0.8\% & 0.0\% & 0.0\% \\
& Receiver 4* & 99.4\%  & 0.6\% & 0.6\% & 0.0\% & 0.0\% \\
\SetCell[r=4]{l} CW Jamming & Receiver 1 & 0.2\% & 99.8\% & 99.0\% & 0.0\% & 0.8\% \\
& Receiver 2 & 0.3\% & 99.7\% & 99.6\% & 0.0\% & 0.1\% \\
& Receiver 3* & 0.2\% & 99.8\% & 99.7\% & 0.0\% & 0.1\% \\
& Receiver 4* & 0.0\%        & 100\% & 100\% &  0.0\% & 0.0\% \\
\SetCell[r=4]{l} Sweep Jamming & Receiver 1 & 0.4\% & 99.6\% & 99.2\% & 0.0\% & 0.4\% \\
& Receiver 2 & 0.5\% & 99.5\% & 99.4\% & 0.0\% & 0.1\%\\
& Receiver 3 & 0.2\% & 99.8\% & 95.8\% & 0.0\% & 4.0\%\\
& Receiver 4 & 0.2\%       & 50.0\% & 95.4\% & 49.8\% & 4.4\% \\
\SetCell[r=4]{l} PRN Jamming & Receiver 1 & 0.2\% & 99.8\% & 95.8\% & 0.0\% & 4.0\% \\
& Receiver 2 & 0.3\% & 86.9\% & 73.9\% & 12.8\% & 25.8\% \\
& Receiver 3 & 0.2\% & 99.8\% & 94.9\% & 0.0\% & 4.9\% \\
& Receiver 4 & 0.1\% & 49.5\% & 94.4\% & 50.4\% & 5.5\%\\
\end{tblr}

\vspace{2mm}
\footnotesize{\textit{*} Receiver recording started at 12:22:00 UTC}
\end{table}

For interference classification using SBAS signals, accuracy is approximately 99\% across all three jamming types, with two exceptions. Receiver 4 under sweep and PRN jamming, where accuracy drops to 50\% due to false spoofing classifications, and Receiver 2 under PRN jamming, where accuracy drops to 87\%. When classification is instead performed using GPS satellites, accuracy remains at 99\% for CW jamming but decreases to 95\% for sweep and PRN jamming. As with SBAS-based classification, Receiver 2 again shows reduced accuracy under PRN jamming. Because GPS-based classification uses all in-view satellites, individual cases in which one or two satellites are misclassified have limited effect on overall accuracy, since the majority of satellites are correctly classified. This also explains why GPS-based classification outperforms SBAS-based classification for Receiver 4 under sweep and PRN jamming, as SBAS classification relies on a single satellite and is therefore more sensitive to individual misclassifications. Although Figure~\ref{fig:Nav25_jam1_gps} appears to show multiple spoofing misclassifications, closer inspection reveals that only a small subset of GPS satellites are affected.

These classification errors primarily occur under PRN jamming, when an increase in received power is not matched by a corresponding degradation in C/N\textsubscript{0} at some receivers, producing false spoofing indications. This mismatch is driven in part by the limited spectral resolution of the received power metric, which is constrained to 500 kHz. Overall, these results demonstrate the effectiveness of the proposed RFI monitoring system for detecting and characterizing jamming, while also highlighting its limitations. The classification errors observed at individual receivers could be mitigated by deploying multiple receivers in a network, which would allow misclassifications to be identified and isolated through cross-receiver comparison.

Compared to existing COTS-based RFI monitoring methods, the proposed approach achieves improved RFI detection accuracy, primarily because it uses received power rather than C/N\textsubscript{0}. This makes detection insensitive to line-of-sight blockages and other environmental effects that degrade C/N\textsubscript{0}. Furthermore, unlike AGC-based methods, the proposed approach can distinguish interference in the GPS L1 band from interference in adjacent GLONASS and BeiDou bands. AGC-based methods are further limited for RFI classification because the AGC saturates during most high-power interference events, providing little information on jamming strength to compare against measured C/N\textsubscript{0}. In contrast, the proposed received-power metric remains unsaturated and preserves its linear relationship with C/N\textsubscript{0}, enabling effective jamming classification.

\subsection{Jammertest Day 2 - Jamming and Spoofing}

\begin{figure}[htb]
    \centering

    \begin{subfigure}{\linewidth}
        \centering
        \includegraphics[scale=0.25]{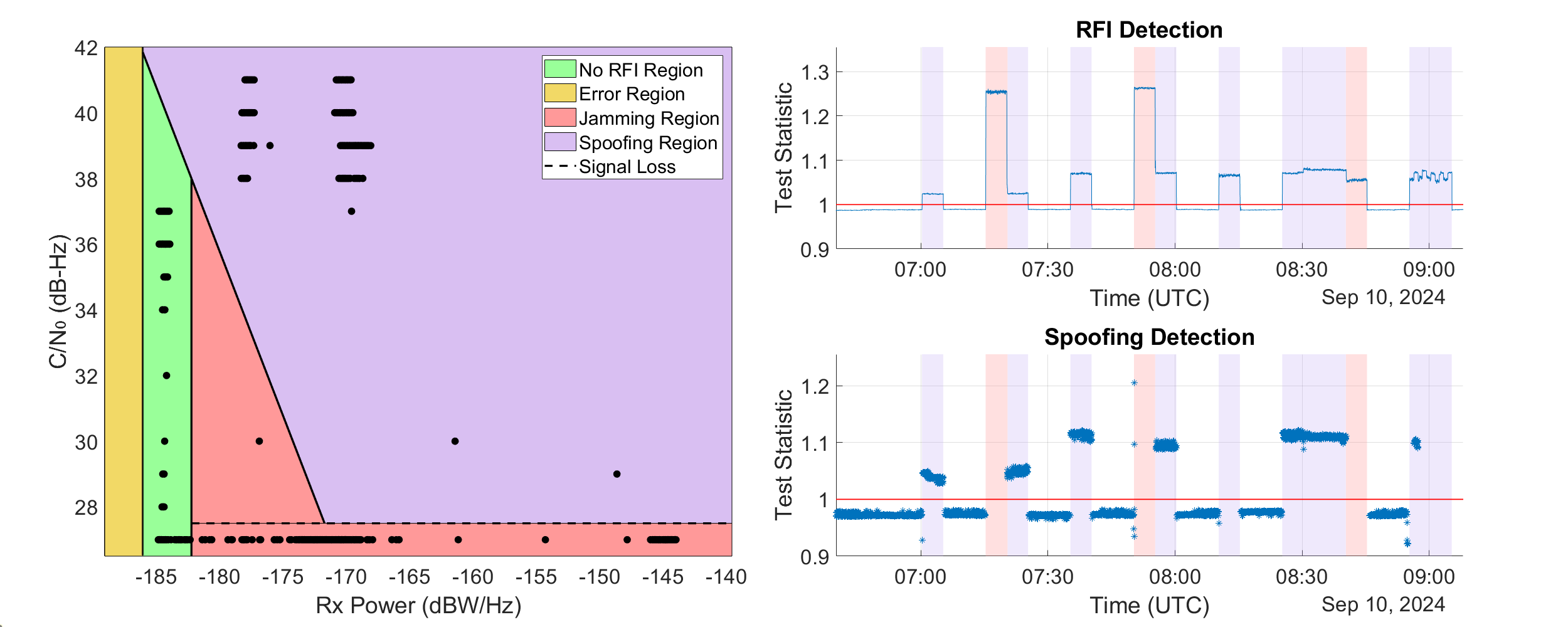}
        \caption{SBAS C/N\textsubscript{0} over received power metric}
        \label{fig:Nav25_jam2_sbas}
    \end{subfigure}

    \vspace{0.3em}

    \begin{subfigure}{\linewidth}
        \centering
        \includegraphics[scale=0.25]{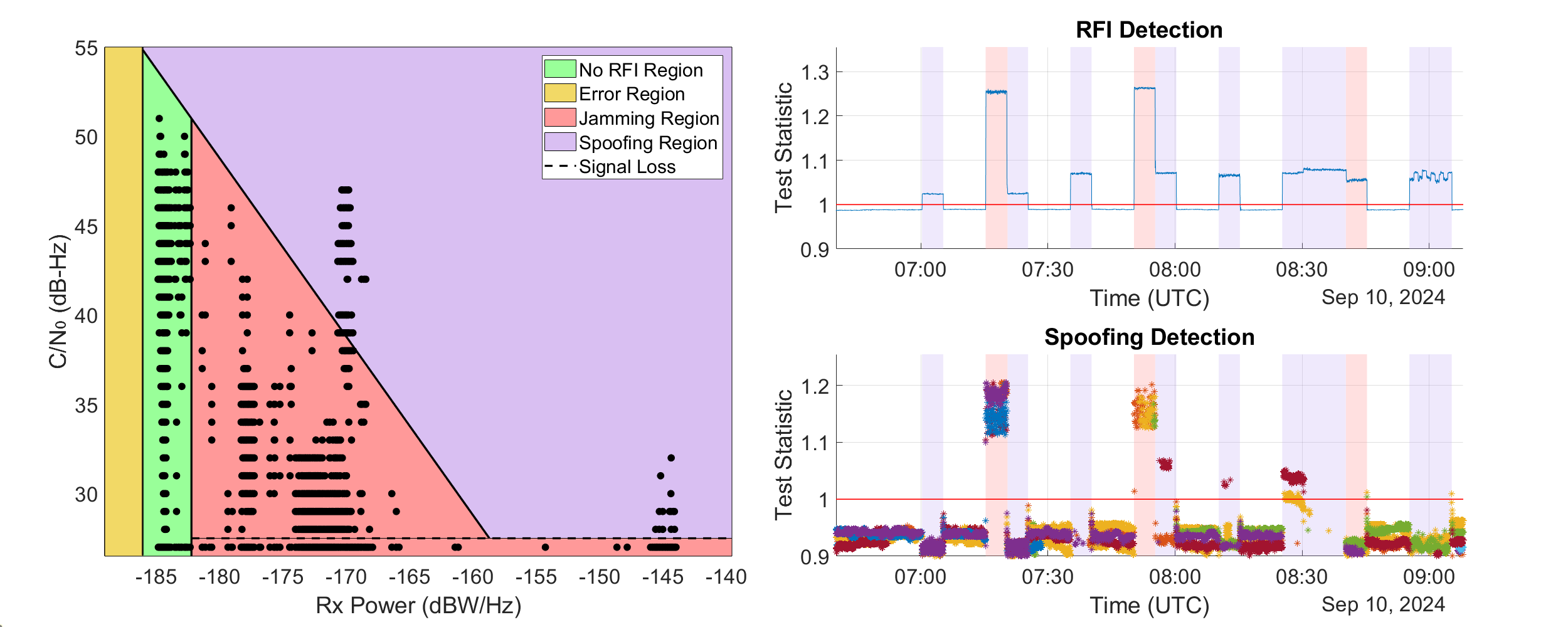}
        \caption{GPS C/N\textsubscript{0} over received power metric}
        \label{fig:Nav25_jam2_gps}
    \end{subfigure}

    \caption{C/N\textsubscript{0} over received power metric from Day 2 of Jammertest (Receiver 3)}
    \label{fig:Nav25_Jam_Day2}
\end{figure}

Figure~\ref{fig:Nav25_Jam_Day2} shows results from one of the three receivers deployed on 10 September, selected as a representative example. Results for the remaining receivers are summarized in Table~\ref{tab:jam_test_matrix_spoof}. Day 2 includes both jamming and spoofing tests, based on a two-and-a-half-hour recording collected that day. Jamming periods are highlighted in red and spoofing periods in purple on the time-varying plots to indicate ground truth.  

As with the Day 1 results, Table~\ref{tab:jam_test_matrix_spoof} summarizes the RFI flag results reported by each receiver for the different RFI scenarios, with ground-truth cells shaded green. The RFI detection accuracy shown in the "No RFI Flag" column exceeds 99.5\%, consistent with the Day 1 results. For the jamming portion of the tests, classification accuracy is also consistent with Day 1, ranging from 95\% to 99\% when using either GPS or SBAS satellites. For spoofing classification, however, accuracy is substantially lower: 53\% to 67\% using SBAS satellites and nearly 0\% using GPS satellites.

\begin{table}[!htbp]
\caption{Jammertest Day 2 (Spoofing) resulting RFI flags, with ground truth highlighted in green}
\label{tab:jam_test_matrix_spoof}
\begin{tblr}{
  colspec = {X[2.2,l] X[1,c] X[1,c] X[1,c] X[1,c] X[1,c] X[1,c]},
  width = \textwidth,
  row{1-2} = {bg=black!20, font=\bfseries\small, valign=m}, 
  row{3-5}   = {bg=black!10, font=\small}, 
  row{6-8}   = {white, font=\small},       
  row{9-11}  = {bg=black!10, font=\small}, 
  cell{3-5}{3}   = {bg=green!15},          
  cell{3-5}{4-7} = {bg=orange!15},         
  cell{6-8}{4-5} = {bg=green!15},         
  cell{6-8}{3}   = {bg=orange!15},        
  cell{6-8}{6-7} = {bg=orange!15},        
  cell{9-11}{6-7} = {bg=green!15},         
  cell{9-11}{3-5}   = {bg=orange!15},        
  hline{Z} = {1pt, solid, black!60},
  rowsep = 0.75pt
}
\SetCell[r=2]{l} RFI Scenario & \SetCell[r=2]{c} Receiver & \SetCell[r=2]{c} No RFI Flag & \SetCell[c=2]{c} Jamming Flag & & \SetCell[c=2]{c} Spoofing Flag & \\
& & & SBAS & GPS & SBAS & GPS \\ 
\SetCell[r=3]{l} No Testing & Receiver 1 & 99.9\% & 0.1\% & 0.1\% & 0.0\% & 0.0\% \\
& Receiver 2 & 99.7\% & 0.2\% & 0.3\% & 0.1\% & 0.0\% \\
& Receiver 3 & 99.8\% & 0.2\% & 0.2\% & 0.0\% & 0.0\% \\
\SetCell[r=3]{l} Jamming & Receiver 1 & 0.0\% & 99.8\% & 94.6\% & 0.2\% & 5.4\% \\
& Receiver 2 & 0.0\% & 97.6\% & 99.7\% & 2.4\% & 0.3\% \\
& Receiver 3 & 0.2\% & 99.1\% & 99.6\% & 0.7\% & 0.2\% \\
\SetCell[r=3]{l} Spoofing (Meacon) & Receiver 1 & 0.2\% & 46.5\% & 99.8\% & 53.3\% & 0.0\% \\
& Receiver 2 & 0.1\% & 34.6\% & 99.8\% & 65.3\% & 0.1\% \\
& Receiver 3 & 0.3\% & 33.0\% & 96.8\% & 66.7\% & 2.9\% \\
\end{tblr}
\end{table}

For SBAS-based classification, the false classifications are not caused by the proposed methodology itself but by the receiver failing to track the spoofed signals. As shown in Figure~\ref{fig:Nav25_jam2_sbas}, the spoofing test statistic correctly identifies spoofing in five of the seven test scenarios. In the remaining two, no spoofed signals were tracked by the receiver, preventing classification. The lower accuracy of the GPS-based metric is attributed to two factors, the receiver failing to track spoofed satellites, and the C/N\textsubscript{0} of tracked satellites falling within the range expected under jamming.
Because the nominal C/N\textsubscript{0} variation of GPS satellites is much larger than that of SBAS satellites, a spoofed signal must substantially overpower the expected value before spoofing detection can occur. These results highlight a limitation of using GPS signals for spoofing classification. While GPS signals can serve as a useful reference, their higher C/N\textsubscript{0} variability makes them less reliable than SBAS signals for distinguishing spoofing from jamming. 

The acquisition and tracking of spoofed signals reflects a broader constraint of receiver-based spoofing detection. Absent overpowering the authentic signal, jamming it prior to spoofing, using unused PRNs, or performing a slow spoofing lift-off attack, a conventional tracking loop will typically continue tracking the authentic signal, and the receiver will show no indication of spoofing. Identifying a spoofed signal that coexists with the authentic one generally requires additional correlators to search for extra signal copies beyond those already being tracked. Such a feature would require modification of the COTS hardware, highlighting a limitation of relying exclusively on COTS components. 

Compared to existing COTS-based RFI monitoring methods, spoofing detection would not be possible using an AGC-based approach, since the AGC saturates under high-power interference and cannot provide a reference value for RFI level. Furthermore, during the spoofing tests, only a limited number of GPS satellites remained tracked by the receiver. Without SBAS signals, identifying the spoofing trend would therefore be challenging. It should be noted that the spoofing scenarios presented here consist of meaconing, in which all GNSS signals are rebroadcast. In scenarios involving GPS-only spoofing, GPS signals would instead be required for spoofing detection.

\section{Discussion and Future Work}

The proposed RFI monitoring methodology achieves the primary objectives of this work. First, it demonstrates that effective detection and characterization of GNSS interference can be achieved using low-cost COTS receivers. Second, by leveraging calibrated received power instead of relying solely on C/N\textsubscript{0} or AGC, the proposed approach provides a significant improvement in RFI detection accuracy while enabling the ability to distinguish between jamming and spoofing. Finally, the automated calibration procedure minimizes manual intervention, enabling straightforward deployment at new locations with minimal setup effort.

The evaluation also highlights several limitations of relying solely on COTS receiver platforms for RFI monitoring. Spoofing classification is only possible once the receiver has acquired the spoofed signal, preventing classification when spoofed signals are not tracked. In addition, false spoofing indications are observed for certain receiver locations and interference types, particularly PRN jamming, where the degradation in C/N\textsubscript{0} does not always follow the expected relationship with the measured received power. This behavior is likely caused by the limited 500 kHz resolution of the receiver FFT used to estimate received power, which does not capture the PRN structure of interference. Finally, while GPS-based spoofing classification remains valuable for identifying GPS-only spoofing events, its performance is limited by the larger variability of GPS C/N\textsubscript{0} compared with SBAS signals.

Several approaches can reduce the impact of these limitations. Improving spoofing signal acquisition would require modified receiver firmware capable of acquiring spoofed signals more reliably. Nevertheless, even when spoofing cannot be classified, the proposed methodology continues to provide reliable RFI detection, ensuring that interference events are still identified and reported. False spoofing indications can be mitigated by validating the pseudoranges whenever a spoofing flag is raised. Since the monitored receivers are stationary, inconsistent pseudorange measurements provide an independent indication of spoofing and can be used to confirm or reject the power-based classification. Furthermore, deploying receivers as a network enables observations from multiple receivers to be combined, reducing the impact of individual receiver misclassifications caused by poor line-of-sight conditions or other local effects.

Future work will focus on extending the proposed methodology beyond single-receiver detection and classification. Nest steps include incorporating pseudorange consistency checks to validate spoofing detections, fusing measurements from multiple receivers to improve system-level detection and classification accuracy, and investigating how the received power measurment can be further used for localization of interference sources. These enhancements aim to enable the proposed monitoring approach to become a comprehensive RFI monitoring and localization system suitable for the protection of critical infrastructure.

\section*{Acknowledgments}
We acknowledge The Aerospace Corporation under their University Partnership Program, the DOT's CARNATIONS center, and the FAA for supporting this effort. The authors also acknowledges the use of Large Language Models (LLM) for editing purposes. 

\nocite{*}
\printbibliography[title=References]

\end{document}